\documentclass[twocolumn,showpacs,preprintnumbers,amsmath,amssymb]{revtex4-1}

\usepackage{dcolumn}
\usepackage{float}
\usepackage{graphicx}
\usepackage{siunitx}

\DeclareMathOperator\erf{erf}

\begin{document}
\title{Highly efficient optical quantum memory with long coherence time in cold atoms}

\author{Y.-W. Cho$^{1}$, G.~T. Campbell$^{1}$, J.~L.~Everett$^{1}$, J. Bernu$^{1}$, D.~B. Higginbottom$^{1}$, M.~T.~Cao$^{1}$, J. Geng$^{1}$, N.~P. Robins$^{2}$, P.~K. Lam$^{1}$ and B.~C. Buchler$^{1*}$}

\affiliation{$^{1}$Centre for Quantum Computation and Communication Technology; $^{2}$Department of Quantum Science, The Australian National University, Canberra, Australia; $^{*}$Corresponding author: ben.buchler@anu.edu.au}

\begin{abstract}
Optical quantum memory is an essential element for long distance quantum communication and photonic quantum computation protocols.  The practical implementation of such protocols requires an efficient quantum memory with long coherence time.  Beating the no-cloning limit, for example, requires efficiencies above 50\%.  An ideal optical fibre loop has a loss of 50\% in 100~$\mu s$, and until now no universal quantum memory has beaten this time-efficiency limit. Here, we report results of a gradient echo memory (GEM) experiment in a cold atomic ensemble with a $1/e$ coherence time up to 1~ms and maximum efficiency up to  $87\pm2\%$ for short storage times.  Our experimental data demonstrates greater than 50\% efficiency for storage times up to 0.6~ms.  Quantum storage ability is verified beyond the ideal fibre limit using heterodyne tomography of small coherent states.
\end{abstract}
\maketitle

\section{Introduction}
A universal optical quantum memory can store an unknown input state of light and release it on demand with high efficiency and without added noise.  The development of such memories is likely to be integral to the development of future quantum information technologies such as the quantum repeater \cite{Lvovsky09,Sangouard:2011bp}, quantum internet \cite{Kimble08} and photonic quantum computation \cite{Knill01}. 

A number of techniques have been developed and explored. Electromagnetically induced transparency (EIT) has been used to show efficiencies up to 78\% \cite{Chen13} and storage lifetimes \cite{lifetime} close to one minute \cite{Heinze:2013dh}. Raman schemes have shown to have very large time-bandwidth products with efficiencies up to 30\% \cite{Reim10,Reim11} and may be particularly suitable for high bandwidth on-demand single photon sources. The atomic frequency comb (AFC)  has shown efficiencies up to 56\%  in a cavity  \cite{Sabooni:2013cr} and on-demand storage lifetimes up to 1~ms \cite{Jobez:2015gt}.  To date the highest efficiency protocol has been the gradient echo memory (GEM) scheme, which has shown 87\% recall in warm vapour \cite{Hosseini11} and 69\% in solid state \cite{Hedges10}. The storage lifetimes in these experiments were on the order of 10~$\mu s$. Laser cooled atoms have yielded a GEM efficiency of 80\% \cite{Sparkes13} combined with a lifetime of about 120~$\mu s$.

When considering the efficiency of a quantum memory there are numerous possible benchmarks depending on the application. One important threshold is $50\%$ total storage and recall efficiency.  In the absence of added noise, a quantum memory performing above this level can beat the no-cloning limit without post-selection \cite{Grosshans01}. It is also a limit that must be surpassed for error correction in some photonic quantum computing protocols \cite{Varnava06}. Although recent experimental demonstrations have shown efficiency of greater than $50\%$ in different systems, so far the coherence time at $50\%$ efficiency is limited to tens of $\mu$s \cite{Hedges10, Hosseini11a, Hosseini11, Riedl:2012gv,Chen13, Sparkes13, Sabooni:2013cr,Afzelius:2014ua}. Achieving high storage efficiency and long coherence time simultaneously is still a challenging goal.

In this work, we present results of a GEM experiment in an ultra-high optical depth (OD) ensemble of laser cooled $^{87}$Rb atoms. The storage efficiency reaches $87\%$ for storage times on the order of the pulse width. The memory lifetime is 1~ms, allowing us to recall at an efficiency above  50\% for times up to 0.6~ms. This is six times longer than could be achieved in an ideal fibre loop. To show that this system is also capable of preserving quantum states, we use optical heterodyne tomography to map out the added noise, showing that our system beats the quantum no-cloning limit.

\section{Preparation of cold atoms}
Our $^{87}$Rb atoms are cooled in a magneto-optical trap (MOT), as shown in Fig.~\ref{fig1} (a). To obtain a large optical depth (OD), our cold atomic medium has an elongated shape. We use rectangular (2D) coils to produce a 2-dimensional quadrupole field for radial confinement and capping-coils that provide axial confinement near the edges of the ensemble \cite{Sparkes13, Geng14}. The resulting cloud contains around $10^{10}$ atoms and has a length of 5~cm. Three pairs of coils mounted around the optical table are used to compensate static magnetic fields. During the loading phase, the trapping (or cooling) laser is red detuned by 35~MHz from the D2 $F=2 \rightarrow F'=3$ transition, while the repumping laser is resonant with the D2 $F=1\rightarrow F'=2$ transition.

After a 170~ms loading phase, we employ a temporal dark spontaneous-force optical trap technique \cite{Petrich94} to compress the MOT radially, thereby increasing the atomic density. We use the 2D MOT coils to smoothly ramp up the magnetic field gradient in the radial direction over a period of 20~ms while red-detuning the repumping laser by 25~MHz. The 2D MOT and axial coils are then switched off and the cooling laser is further detuned by 63~MHz to apply polarisation gradient cooling for 1.3~ms. Because our memory relies on one particular Zeeman coherence, it is favourable to pump atoms into a single Zeeman sublevel  \cite{Hsiao14}.  Zeeman optical pumping is therefore employed for 470~$\mu s$ after the polarisation gradient cooling phase to pump atoms into the $m_F=+1$ Zeeman state. 

The light storage experiments were done with all optical and magnetic MOT fields switched off. The decay of the magnetic trapping fields and eddy currents induced in nearby metallic objects was found to have an exponential time constant of $\sim 0.7$~ms.  The probe-field to be stored in the atoms was injected 2.2~ms after the coils were switched off to allow sufficient decay of magnetic field fluctuations.  The typical OD at this time is $\approx$600 on the  D1 $F=1 \rightarrow F'=2$ transition.  

\section{Implementation of the GEM scheme}
The GEM scheme \cite{Hosseini11a, Hosseini11} uses an atomic medium  that is inhomogeneously broadened by a magnetic field gradient.  Light is absorbed into the atomic ensemble to create an atomic coherence, which then dephases due to the applied broadening gradient. The reversal of the gradient gives rise to the rephasing of the atomic coherence and thus recall of the stored light.

\begin{figure}[t]
\centering
\includegraphics[width=3.4 in]{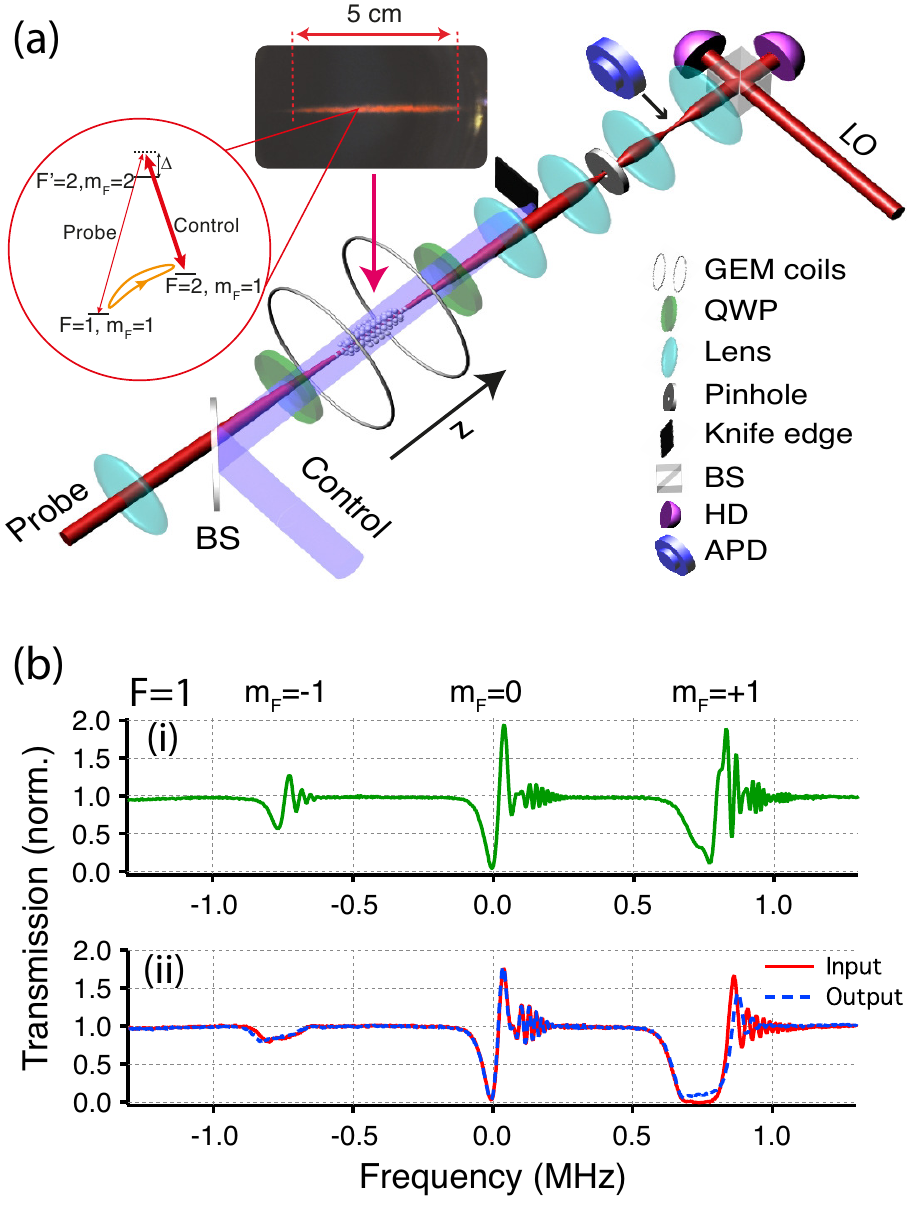}
\caption{(a) Experimental schematic.  A cold  $^{87}$Rb cloud is prepared in the MOT.  The photo of the atomic cloud was taken using a 30s exposure. The probe and control-fields are combined with angle $\theta$ at a beam splitter (BS). The polarisations of probe and control fields are set to be the same circular by a quarter-wave plate (QWP). The probe field  is measured either by an avalanche photodiode (APD) or by a heterodyne detection (HD) system with a local oscillator (LO) beam. The GEM coils generate a magnetic field in the $z$-direction with a reversible gradient. The atomic level scheme illustrates the relationship between the probe and control-fields. (b) Measured Raman absorption spectra: (i) unbroadened Raman lines, (ii) Raman lines broadened by the magnetic field gradients used during the input and output stages. The oscillations are due to free-induction decay.
}
\label{fig1}
\end{figure}

Figure~\ref{fig1} (a) illustrates the experimental setup of the memory system. Both the control and probe-fields are derived from a Ti:Sapphire laser (not shown), where the frequencies and the amplitudes are independently controlled by acousto-optic modulators and an electro-optic modulator \cite{Sparkes13}. The probe-field is focused at the centre of the MOT with a beam waist of 110~$\mu$m. The collimated control-field has a large beam waist of $\approx$10~mm to cover the interaction region with a near homogenous intensity. Spatial filtering is used to separate the probe light from the much more powerful control field. The probe is focussed through a 100~$\mu$m diameter pinhole and a knife-edge is used when there is an angle, $\theta$, between the beams. We achieve $>$34~dB attenuation of the control-field while allowing $>$92\% transmission for the probe. The probe-field is then measured either by an avalanche photodiode (APD) or a balanced heterodyne detection system.
 
 The GEM coils are used to create a magnetic field with a z-component that has a near-uniform gradient in the $z$-direction.  This creates a controllable spatial atomic frequency detuning, $\delta(z,t)$, that is close to linear in $z$. This gradient can be reversed to activate the recall of the stored light \cite{Hosseini11}. Using only two coils allows adequate optical access, although this does affect the homogeneity of the atomic frequency gradient.  This can impact the memory decay rate, as we will describe in section \ref{higheff}.
 
 As shown by the level scheme in Fig.~\ref{fig1} (a), the probe and control beams are used to form a Raman absorption line. It is this line that is used to absorb the probe light.  As in previous GEM experiments using three-level atoms, the mapping of the probe into and out of the memory is then mediated by the control-field and the applied magnetic field gradient \cite{Hosseini11}.
 
\section{Raman line measurement}
We consider the off-resonant Raman lines in the $^{87}$Rb D1 line using the level scheme shown in Fig.~\ref{fig1}(a). Both the control and probe fields were blue detuned by $\Delta$=325 MHz from the excited state $F'=2$ and set to be have the same $\sigma^+$ polarisations. The Zeeman degeneracy was lifted by a constant uniform bias field of 0.5~Gauss, which was generated using the GEM coils. We took data for various control laser powers and fitted the measured spectrum to an analytic model. This allowed us to calibrate the control laser power to the square of the Rabi frequency $|\Omega_c|^2$ and OD for each Raman transition. Figure~\ref{fig1} (b) (i) shows representative Raman line data with a uniform magnetic field. We find ODs of 6.3, 38, and 488 for $m_F=-1,0,1$, respectively, demonstrating the efficacy of the afore mentioned optical pumping into the $m_F=1$ level. Applying differing currents through the GEM coils leads to a magnetic field gradient and broadened Raman lines, as shown in Fig.~\ref{fig1}(b). The broadened widths of the input and output gradients were found to be 197 kHz and 210 kHz, respectively.

\section{High efficiency storage} \label{higheff}
A probe pulse with a Gaussian full-width-half-maximum (FWHM) of 6.66~$\mu$s was coherently absorbed into the atoms, and then recalled by changing the sign of the magnetic field gradient. 
\begin{figure}[b!]
\centering
\includegraphics[width=3.4 in]{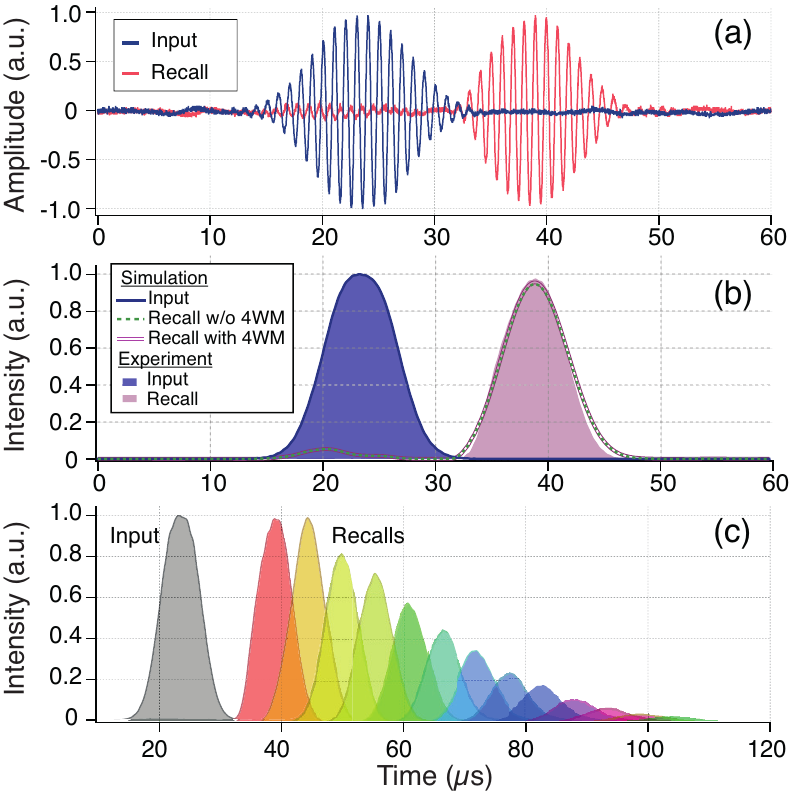}
\caption{Demonstration of high efficiency storage. (a) Experimental heterodyne data for input/recall pulses. The fringe visibility between the local oscillator and probe beams was $>$97\%. (b) Demodulated experimental data (filled regions), averaged over 15 traces, shows an efficiency of  $87\pm2 \%$.  Numerical results with and without 4WM are overlaid. (c) Storage and recall data for different storage times. The decay is relatively quick due to the presence of control and gradient fields, as described in the text.
}
\label{fig2}
\end{figure}
Figure~\ref{fig2} (a) shows the balanced heterodyne signal for input and recall, where the input reference signal was obtained without the MOT. Fig.~\ref{fig2} (b) shows the corresponding demodulated signals averaged over 15 pulses. By directly integrating the area below the input and output pulses, we estimate the efficiency in this experiment to be $87\pm2\%$.  Efficiency measurements using an APD were within experimental uncertainty of the heterodyne results. The APD, however, could only be used when the angle, $\theta$, between the control and probe beams was $\geqslant 0.2^\circ$  so that adequate spatial filtering could be employed to avoid saturating the APD. 

The output pulses experience only slight distortion as a result of storage.  For a single shot, the amplitude overlap of  the output with a Gaussian reference is ($99.3\pm0.1$)\%. We observe neither a frequency shift nor a chirp, both of which are eliminated by adjustment of the magnetic field gradient at recall. A shot-to-shot phase drift of 5$^\circ$ between the input and recalled pulses slightly reduces the mean amplitude overlap to ($99.1 \pm 0.3$)\%.

Four wave mixing (4WM) can lead to noise and amplification in coherent atom-light systems \cite{Geng14}. In order to estimate the impact of 4WM in our experiment we numerically solve the Maxwell-Bloch equations with experimental parameters.  These simulation results \cite{supple} are compared to the experimental data in Fig.~\ref{fig2}(b). The numerical simulation predicts a recall efficiency of 89\% which is within experimental uncertainty of our measured value. Comparing simulations with and without 4WM shows that the power contained in the idler is expected to be 0.9\% of the probe input power, leading to gain in the probe output of 0.9\%. This result is consistent with previous work in warm vapour which found little additional noise in the GEM protocol with similar storage times and efficiencies \cite{Hosseini11a}.

\begin{figure}[t]
\centering
\includegraphics[width=3.4 in]{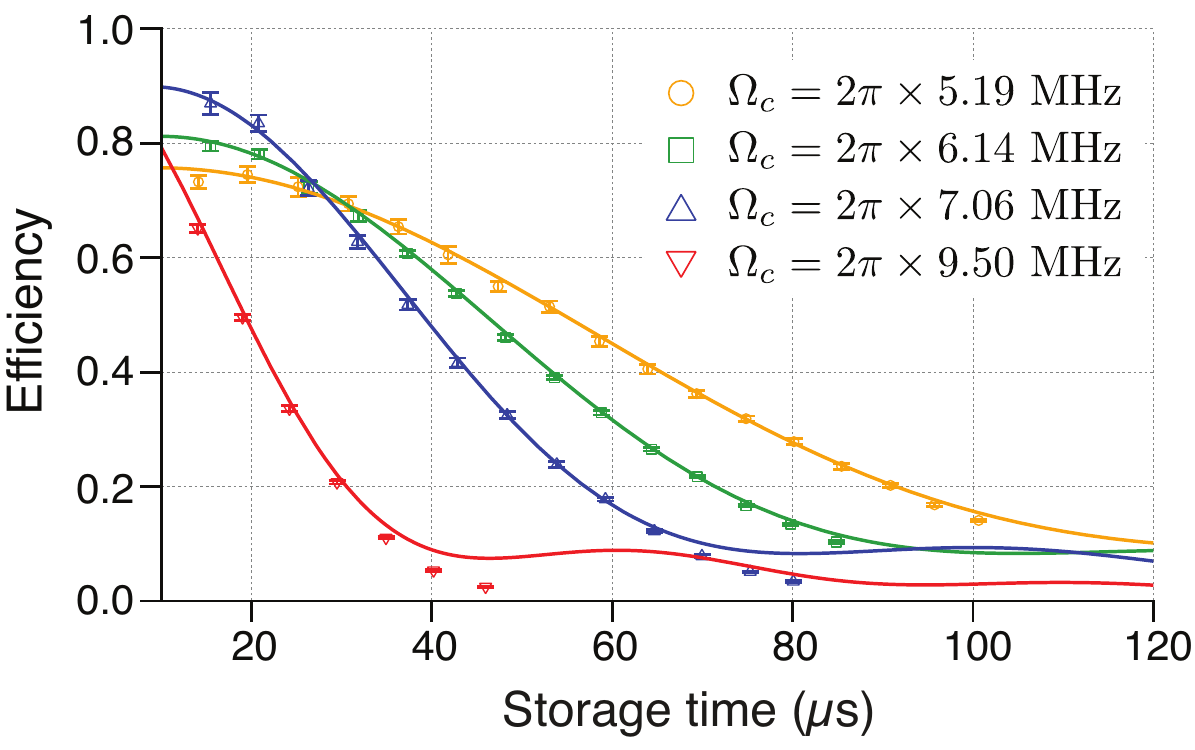}
\caption{Efficiency decay when both the control-field and the magnetic field gradients are left on during storage, shown for varying values of the control-field Rabi frequency $\Omega_c$. Solid lines are fitting curves with a decay model described in Eq.~(\ref{shortdecaymodel}).} 
\label{short}
\end{figure}

Figure~\ref{fig2} (c) shows results for different storage times, obtained by changing the gradient field switching time. For this data, both the control-field and the magnetic gradient were left on during storage. Both these fields have an impact on the storage lifetime. An exponential decay with a rate $\Gamma_{\rm{sc}}/2\pi \simeq \Gamma (\Omega_c/2\Delta)^2 /2\pi$ is expected due to the control-field induced scattering. The magnetic gradient, on the other hand, was observed to give non-exponential decay. We attribute this to the geometry of the GEM coils. We modelled the atomic detuning, $\delta(z)$, induced by the coils up to second order in $z$ by assuming $\delta(z)=\delta_0+\eta^{(1)} z + \eta^{(2)}z^2$.  The constant offset $\delta_0$ lifts the degeneracy of the $m_F$ states. In our experiment, swapping the current in the coils leaves $\delta_0$ unchanged, inverts the sign of $\eta^{(1)}$, but does not invert the sign of $\eta^{(2)}$. Swapping the currents, therefore, does not give full spatial inversion of the field gradient when there is a non-zero $\eta^{(2)}$. From here we derived a magnetic field dependent model of decay assuming a Gaussian input pulse \cite{supple}:
\begin{equation}
\mathcal{E}(t)= \frac{\mathcal{E}_0  \left| \erf \left[ \sqrt{1-i \zeta (t-t_0)}/(\frac{4\sigma}{L})\right]\right|^2}{\sqrt{\zeta^2 (t-t_0)^2 + 1}}e^{-\Gamma_{\rm{sc}} t},
\label{shortdecaymodel}
\end{equation}
where $\zeta \equiv 4 \eta^{(2)} \sigma^2$, $\sigma$ is the spatial width of the spin-wave envelope, the initial efficiency is normalised by $\mathcal{E}_0$ and  $t_0$ is an offset to compensate for the read and write durations. This model shows good agreement with the measured data, as shown in Fig.~\ref{short}.

With the control and gradient fields on, GEM is an effective memory for multiple temporal modes. The result for $\Omega_c=2\pi \times 5.19$ MHz shows a fractional delay of 8.1 at 50\% efficiency, where the fractional delay is defined as the ratio of storage time to FWHM pulse width. We have also directly demonstrated the multimode capacity by storing and recalling a pulse train of 20 Gaussian pulses with an average efficiency of 14\% \cite{supple}. Large multimode capacity and fractional delay is important for time-bin qubit storage and multiplexing \cite{bussieres2013prospective}. 

\section{Long-lifetime single mode storage}
\begin{figure}[t]
\centering
\includegraphics[width=3.4 in]{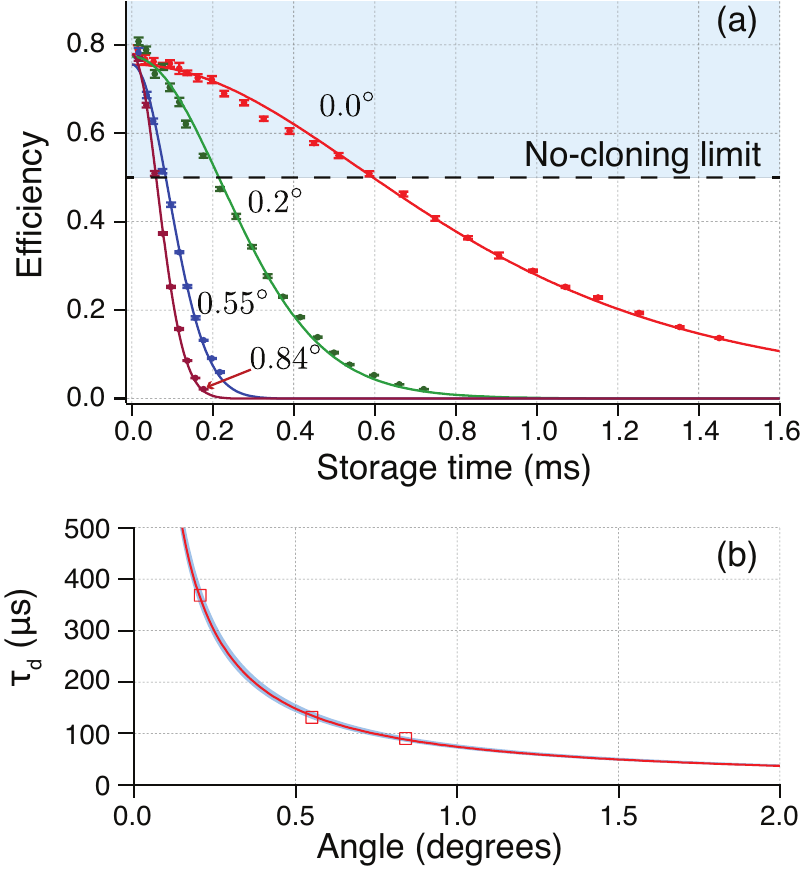}
\caption{The memory lifetime is extended by turning off both the control and the magnetic gradient field. (a) Measured efficiency as a function of storage time. For $\theta=0.0^\circ$, the coherence time is limited by the atom loss, see text for details. The corresponding $e^{-1}$ decay time is 1 ms. Solid lines are the theoretical fit with Eq.~(\ref{decaymodel}). For $\theta=0.2^\circ,~0.55^\circ,$ and $0.84^\circ$, the coherence times are limited by the longitudinal dephasing. (b) Measured coherence time due to the longitudinal dephasing as a function of the angle $\theta$. Solid line shows the calculated curve for $T=100~\mu$K.  The shaded region shows the bounds for fits at $T=110~\mu$K (lower bound) and $T=90~\mu$K (upper bound). 
}
\label{fig4}
\end{figure}

To achieve the best memory lifetime we are free to turn off both the control-field and the gradient magnetic field during the storage duration. Note that the uniform bias magnetic field is still present to mitigate the impact of ambient magnetic field fluctuations which would otherwise severely limit the memory lifetime. Figure~\ref{fig4} shows the experimental results. For the co-propagating case ($\theta=0^{\circ}$), we obtained a memory lifetime of 1 ms. Furthermore, we observed storage with efficiency above $50\%$ for times up to 0.6~ms with a fractional delay of 84.

In the absence of the control-field, the decay in memory efficiency is due now to other mechanisms.  As shown in Fig.~\ref{fig4}, there is a strong dependence of the memory lifetime on the angle between probe and control beam. This is due to the thermal motion of the atoms. For a cold atomic ensemble at temperature $T$, the mean atomic velocity is $\bar{v}=\sqrt{k_B T/m}$, where $k_B$ and $m$ are the Boltzmann constant and the atomic mass, respectively.

Both the radial and longitudinal motion will contribute to the decay of the memory. The radial motion leads to atom loss out of region that contributes to the detected output mode of the memory.  The radial position of an atom can be modelled as $\rho(t)=\sqrt{\rho(0)^2+2\bar{v}^2 t^2}$ to give a characteristic time for the atom loss of $\tau_l=w_0/\bar{v}$, where $w_0$ is the waist of the detected mode.

The longitudinal motion is more complicated to analyse. If $\bar{v}t \gg \lambda_{\rm{SW}}=2\pi/|\bf{k_{{SW}}}|$, where $\bf{k_{{SW}}}$ is the spatial frequency of the GEM spinwave, the atomic coherence will be washed out. The characteristic time for the longitudinal dephasing is given as $\tau_d=\lambda_{\rm{SW}}/2\pi \bar{v}$. Then, the overall efficiency is found to be \cite{Jenkins12}
\begin{equation}
\mathcal{E}(t)=\frac{\mathcal{E}_0}{\left[1+(t/\tau_l)^2\right]^2}\exp{\left[\frac{-(t/\tau_d)^2}{1+(t/\tau_l)^2}\right]}.
\label{decaymodel}
\end{equation}

Unlike the EIT polariton, the GEM polariton propagates in $k$-space at a velocity given by the frequency gradient $\eta^{(1)}$ \cite{Hetet08, Luo13}, i.e. $\mathbf{k_{SW}}{\rm{(t)}}\cdot \hat{\rm{z}}=\mathbf{k_{SW}^{0}}\cdot \hat{\rm{z}}+\eta^{(1)}\rm{(t) t}$.  The initial spatial frequency is given by $\mathbf{k_{SW}^{0}}=\mathbf{k_p}-\mathbf{k_c}$ where $\mathbf{k_p}$ and $\mathbf{k_c}$ are the k-vectors of the probe and control fields, respectively. This initial spatial frequency will be increased for any $\theta\neq 0$. For a given atomic temperature, longitudinal dephasing can be minimised by keeping $|\mathbf{k_{SW}}{\rm{(t)}}|$ small.  This means minimising $\theta$ and leaving the magnetic gradient off where possible, not only to reduce the impact the quadratic gradient term discussed above, but also to keep the wavelength of the spinwave larger.

Given the temperature of our atoms is $T=100~\mu$K and the beam waist is $w_0=110~\mu$m, we estimate  characteristic times of $\tau_l=1.12$ ms and $\tau_d=71$ ms for $\theta=0^\circ$, indicating that we are in a regime dominated by radial diffusion out of the mode of the memory.  Fitting the experimental data at $\theta=0^\circ$ (shown in Fig.~\ref{fig4}a) with $\mathcal{E}(t)\simeq {\mathcal{E}_0}/{\left[1+(t/\tau_l)^2\right]^2}$, we find $\tau_l=1.24$~ms, which is in good agreement with the estimated value.

As expected, the memory lifetime is much reduced as the angle ($\theta$) between control and probe beams is increased. Fixing the longitudinal time constant at $\tau_l=1.24$ ms,  Eq.~(\ref{decaymodel}) can be used to find $\tau_d$ as we increase $\theta$. The data points in Fig.~\ref{fig4}(b) show the angular dependence of $\tau_d$ determined in this way. These points are in excellent agreement with the curve found by assuming $T=100~\mu$K.  An independent determination of the temperature based on measurements of the optical depth as a function of time gave $95~\mu$K. Based on this model, we expect that further improvement is possible by increasing the probe beam size and lowering the temperature. For example, we expect $\tau_l= 6.3$ ms with $T=10~\mu$K and $w_0=220~\mu$m. Eventually, given the horizontal geometry of our experiment, we would become limited by the free-falling atoms leaving the memory mode \cite{Bao12}.

\section{Analysis of the quantum performance of the memory}

To directly verify that the memory can store and recall quantum states without introducing significant decoherence, we performed balanced heterodyne measurements of weak coherent states before and after storage in the memory.  The quantum state of the probe-field was measured by increasing the local oscillator power such that the shot noise was above technical noise within the detection frequency band of ($3.0 \pm 0.1$) MHz. The detected heterodyne signal was then demodulated with sine and cosine functions that were referenced to the phase of a bright pulse. This reference pulse was sent through the memory, with the control-field off, just prior to the input of the probe-field.

\begin{figure}[t]
   \centerline{\includegraphics[width=\columnwidth]{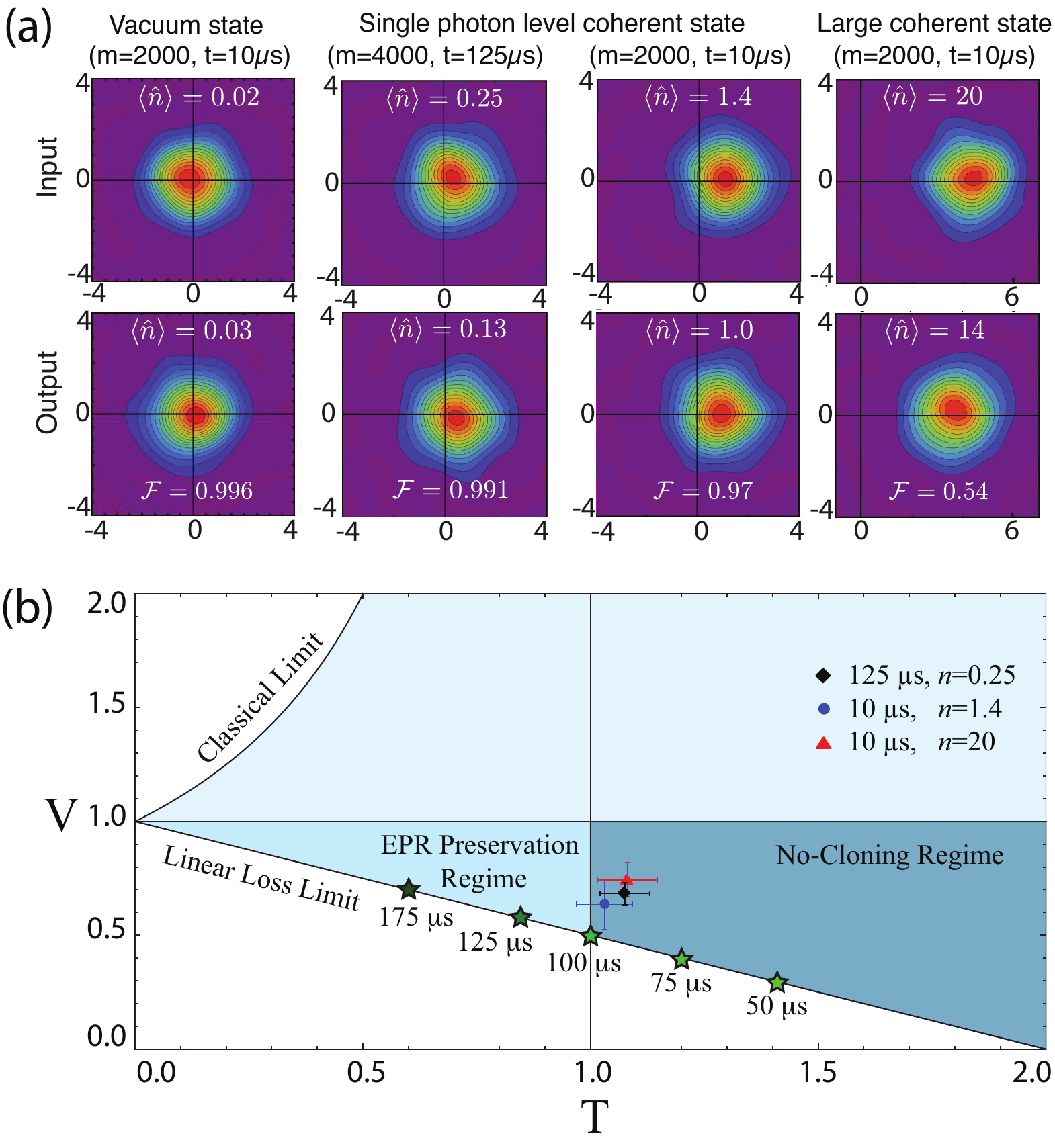}} 
   \caption{(a) Measured Wigner functions of a selection of coherent states at the input and output of the memory.  The parameter `m' is the number of pulses in the dataset, $t$ is the storage time. The mean measured photon numbers are indicated in the plots. The input states (top row) were recorded in the absence of trapped atoms while the output states (bottom row) were recorded after being stored in the memory and recalled. (b) \emph{T-V} characterisation of the quantum performance of the memory. The classical limit assumes the state is measured and recreated with the applicable vacuum penalties for each quadrature.  The linear loss limit is a lower physical bound that an ideal fibre would follow, assuming it adds no noise.  Indicated on this line are some accessible ideal fibre storage times, where 100~$\mu s$ is at the T=1 boundary of the no-cloning limit.}
   \label{wigner}
\end{figure}

The heterodyne detection provides a direct sampling of the Husimi Q-function \cite{Leonhardt199589}, which we estimated from ensembles of heterodyne measurements of individual pulses. The corresponding Wigner functions are then found by performing a deconvolution with the vacuum state. Figure \ref{wigner}(a) shows the measured Wigner functions of the input and output states along with the corresponding photon number and the fidelity between the input and output states. Results show a high recall fidelity (99.6\%) for the vacuum state. This result indicates no significant background noise is present in the memory.  As the photon number increases the phase-space overlap between the input and output is reduced.  This is due to the non-unity efficiency of the memory, which reduces the coherent amplitude of the output relative to the input. The memory efficiency in the heterodyne data was 73\% for 10~$\mu s$ storage and 53\% for 125~$\mu s$ storage.  This is because we were forced to reduce the Raman transition detuning to 160MHz (10~$\mu s$ data) and 200~MHz (125~$\mu s$ data) to enable the laser locking required to maintain stability for the collection of statistical data. (See  \cite{supple} for details.)

To obtain a state-independent characterisation of the quantum performance, we use the \emph{T-V} representation \cite{hetet2008characterization,Ralph98}. The $T$ coefficient is a measure of how well the signal-to-noise ratio of the quantum state is preserved while the $V$ coefficient is a measure of the noise added to the output state.  This allows the definition of three performance regimes, as shown in Fig.~\ref{wigner}(b).  The classical limit is the weakest, followed by the EPR (Einstein-Podolsky-Rosen) preservation regime in the lower left quadrant, and finally the no-cloning regime in the lower right quadrant.  All physical systems are bounded to be above the linear loss limit.  A state of light delayed or stored in an ideal fibre loop, where the only source of state degradation is attenuation, would have \emph{T-V} parameters that track along the linear loss limit.  We have indicated some sample fibre storage times on in the figure, assuming best case fibre attenuation of  0.15~dB/km at 1550~nm.

When calculating the \emph{T-V} parameters, detection losses and noise could obscure noise added by the memory and must be taken into account to avoid overestimating the quantum performance.  The total effective detection efficiency is calculated from the losses due to spatial filtering, heterodyne fringe visibility, the heterodyne detection penalty, the quantum efficiency of our detectors and shot-noise to dark-noise ratio (See \cite{supple} for details).  The 10~$\mu s$ and 125~$\mu s$ storage data had total detection efficiencies of 17\% and 24\% respectively.  Correction for detection efficiency reduces $T$, increases $V$ and also increases the error bounds that appear in our \emph{T-V} data.

The experimental \emph{T-V} points are shown in Fig.~\ref{wigner}(b). We see that they lie above the linear loss limit indicating that there is some additional noise in our system, to which the \emph{T-V} parameterisation is sensitive. Analysis of our heterodyne data indicates this noise is in the phase quadrature of the output states.  We believe that our system is affected by shot-to-shot variation of the MOT parameters, leading to small fluctuations in the recalled phase.  The amount of added noise is not observed to increase as we lengthen the storage time.  For a storage time of 125~$\mu s$ we are more than one standard deviation within the no-cloning regime.  By comparison we see the 125~$\mu s$ point for the ideal  fibre is well outside this regime. It is also worth noting that a real fibre may not achieve ideal noiseless performance due to acoustic fluctuations of the fibre length adding phase noise to states delayed in a fibre.

\section{Comparison of memory systems}
\begin{figure*}[t!]
   \centerline{\includegraphics[width=17.5cm]{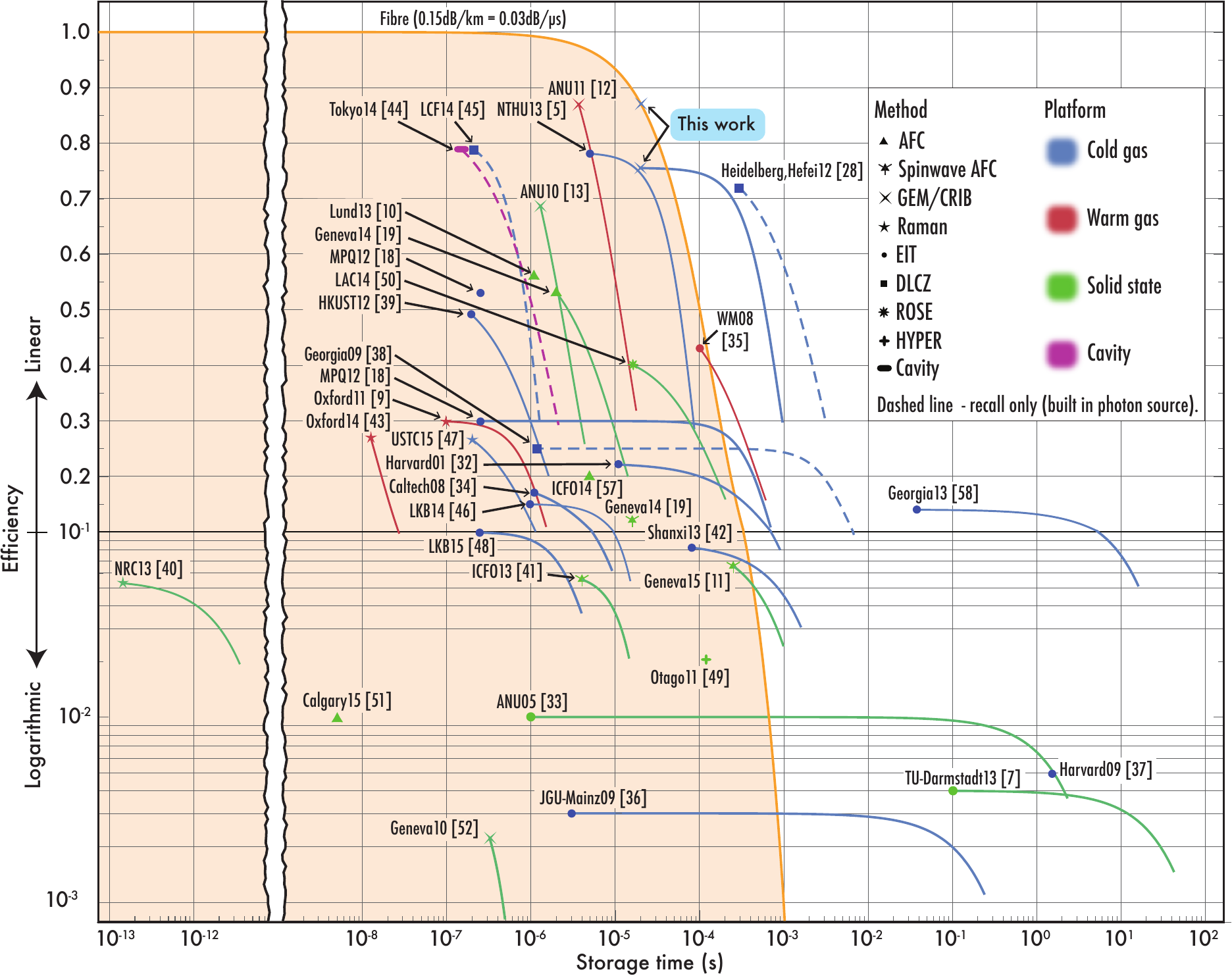}} 
   \caption{Comparison of the efficiency and storage time of quantum memories that output an optical state. Not all the experiments in this graph were done in the quantum regime, but all the techniques are at least theoretically compatible with quantum storage. Both universal \emph{capture-and-release} (solid lines) and \emph{recall-only} (dashed lines) memories are included. Note that preparation efficiencies for \emph{recall-only} memories are not taken into account. The vertical axis of the plot shows measured efficiency; above 0.1 with a linear scale and below 0.1 with a logarithmic scale.  The horizontal axis shows storage time, which spans nearly 15 orders of magnitude. For each experiment a point is plotted on the graph indicating the maximum storage efficiency, which always occurs for the shortest reported storage time.  Where a decay model is given, or data is available, a curve is also plotted showing how the storage efficiency decreases with storage time.  The curves are plotted down to the point where the efficiency reaches $e^{-1}$ of the maximum recorded efficiency.  The yellow curve and shaded region indicate region accessible by an ideal fibre loop at 1550~nm, neglecting losses from input and output coupling. In method CRIB, DLCZ, ROSE and HYPER refer to controlled reversible inhomogeneous broadening, Duan-Lukin-Cirac-Zoller scheme, revival of silenced echo, and hybrid photon echo rephasing, respectively. Published data is labeled by institute, year and citation. Citations that appear in this plot, but not in the text: \cite{Liu:2001wy,Longdell:2005ik,Kimble_nature_sph-QM1,Phillips:PRA:2008,Schnorrberger:2009ev,Zhang:2009fi,Zhao:2009ic,Zhou:2012fw,England:2013uf,deRiedmatten:2013vb,Xu:2013ir,Sprague:2014fd,Yoshikawa:2014wf,Bimbard:2014fy,Nicolas:2014fn,Ding:2015bg,Gouraud:2015je,McAuslan:2011up}}
   \label{comp}
\end{figure*}

 For quantum repeaters in optical fibre networks, the point of the memory is to overcome losses in fibre.  An obvious benchmark for a memory is therefore the performance of optical storage compared to an ideal fibre loop. In Fig.~\ref{comp} we plot the efficiency of our memory as a function of time with the shaded region showing what is accessible by fibre delay. The ideal fibre loses 50\% of the input light after 100~$\mu$s. Our system beats this time by a factor of 6 and is, to the best of our knowledge, the only coherent optical memory that has demonstrated storage above 50\% efficiency for a time that beats delay in an ideal fibre.  

Also shown in Fig.~\ref{comp} is a comparison of various published quantum-compatible memory schemes, restricted to ensemble systems that are capable of giving an optical output state.  Ensemble systems are the most general form of quantum memory as they can store both single and multi-photon quantum states. In this plot we have included data from a range of different published experiments, not all of which provide an analysis of the quantum performance of the experiment, although all the protocols used have been shown to be compatible with quantum storage. We have categorised the experiments according to platform and protocol, as labelled. Most of these experiments work as universal \emph{capture-and-release} memories that accept some travelling mode of light as an input.  A few recent experiments (indicated by dashed lines) make use of photon sources that are built into the memory.  Since in-coupling of light is not required such a system is attractive for building deterministic single photon sources. One must, however, be mindful of the success rate of preparing a photon in the memory.  This preparation efficiency is not included in Fig.~\ref{comp}.  Furthermore, some quantum repeater protocols require universal memories \cite{Sangouard:2011bp}.  It is also worth noting that only \cite{Dajczgewand:2014et, Saglamyurek:2015es,Lauritzen:2010go} work directly at 1550~nm, at the same wavelength as the ideal fibre loop, making these schemes more readily compatible with fibre-optic networks.  All other memories on this plot would require wavelength translation \cite{Tanzilli:2005jd,Ikuta:2011de,Zaske:2011id,Albrecht:2014cf} to work at telecom wavelengths, as demonstrated in \cite{Maring:dp} with 20\% conversion efficiency.

This plot does not explicitly show a third important memory property, namely the time-bandwidth product. For many of the memories shown, however, there is an implicit indication of this property. To get the best efficiency a short storage time is advantageous, which means making the shortest pulse that will fit into the memory and recalling it as soon as possible, i.e. about one pulse-width later.  For this reason, the shortest recorded storage time is, for most experiments, a proxy for the inverse bandwidth of the memory. If the storage time is much longer than the shortest storable pulse width then the decay curve shows a long flat plateau, which is indicative of a memory where the experimental data has shown a high time-bandwidth product.  A notable exception is the Raman memory \cite{Reim11}, which used a pulse width of 300~ps, although the shortest reported storage time was 100~ns.

In terms of memory platform, cold-atom based systems have the distinction of providing both very long storage \cite{Dudin:2013ew} and the highest recorded efficiency, presented in this paper.  It is unlikely that these characteristics can be combined, however, as the long storage times shown in \cite{Dudin:2013ew} made use of an optical lattice, which is not conducive to the high optical depth required for maximising the efficiency.  
Solid state systems appear to hold the best prospects for combining long storage and high optical depth \cite{Zhong15}, although this has yet to be demonstrated in an experiment.

\section{Conclusion}
 Extending the memory lifetime while maintaining high storage efficiency is an important step toward a practical quantum memory. In addition to efficiencies up to 87\%, our system demonstrates millisecond storage lifetime which translates into 0.6~ms of storage above the 50\% no-cloning  threshold. Our results and modelling suggest that our quantum memory exhibits good phase stability as well as negligible added photon noise from residual nonlinear optical process.  The decay mechanisms that limit the storage time have been characterised, and this work points the way to further improvements in storage time.
\vspace{5mm}

\section*{Funding Information}
This work is funded by the Australian Research Council (ARC) Centre of Excellence Program (CE110001027). YWC is partially supported by the National Research Foundation of Korea (2014R1A6A3A03056704). PKL and BCB are supported by ARC  Laureate Fellowship (FL150100019) and Future Fellowship (FT100100048), respectively.
 
\section*{Acknowledgments}
We thank S. M. Assad for advice regarding the quantum data analysis. 

\bibliography{library}

\begin{thebibliography}{59}%
\makeatletter
\providecommand \@ifxundefined [1]{%
 \@ifx{#1\undefined}
}%
\providecommand \@ifnum [1]{%
 \ifnum #1\expandafter \@firstoftwo
 \else \expandafter \@secondoftwo
 \fi
}%
\providecommand \@ifx [1]{%
 \ifx #1\expandafter \@firstoftwo
 \else \expandafter \@secondoftwo
 \fi
}%
\providecommand \natexlab [1]{#1}%
\providecommand \enquote  [1]{``#1''}%
\providecommand \bibnamefont  [1]{#1}%
\providecommand \bibfnamefont [1]{#1}%
\providecommand \citenamefont [1]{#1}%
\providecommand \href@noop [0]{\@secondoftwo}%
\providecommand \href [0]{\begingroup \@sanitize@url \@href}%
\providecommand \@href[1]{\@@startlink{#1}\@@href}%
\providecommand \@@href[1]{\endgroup#1\@@endlink}%
\providecommand \@sanitize@url [0]{\catcode `\\12\catcode `\$12\catcode
  `\&12\catcode `\#12\catcode `\^12\catcode `\_12\catcode `\%12\relax}%
\providecommand \@@startlink[1]{}%
\providecommand \@@endlink[0]{}%
\providecommand \url  [0]{\begingroup\@sanitize@url \@url }%
\providecommand \@url [1]{\endgroup\@href {#1}{\urlprefix }}%
\providecommand \urlprefix  [0]{URL }%
\providecommand \Eprint [0]{\href }%
\providecommand \doibase [0]{http://dx.doi.org/}%
\providecommand \selectlanguage [0]{\@gobble}%
\providecommand \bibinfo  [0]{\@secondoftwo}%
\providecommand \bibfield  [0]{\@secondoftwo}%
\providecommand \translation [1]{[#1]}%
\providecommand \BibitemOpen [0]{}%
\providecommand \bibitemStop [0]{}%
\providecommand \bibitemNoStop [0]{.\EOS\space}%
\providecommand \EOS [0]{\spacefactor3000\relax}%
\providecommand \BibitemShut  [1]{\csname bibitem#1\endcsname}%
\let\auto@bib@innerbib\@empty
\bibitem [{\citenamefont {Lvovsky}\ \emph {et~al.}(2009)\citenamefont
  {Lvovsky}, \citenamefont {Sanders},\ and\ \citenamefont
  {Tittel}}]{Lvovsky09}%
  \BibitemOpen
  \bibfield  {author} {\bibinfo {author} {\bibfnamefont {A.~I.}\ \bibnamefont
  {Lvovsky}}, \bibinfo {author} {\bibfnamefont {B.~C.}\ \bibnamefont
  {Sanders}}, \ and\ \bibinfo {author} {\bibfnamefont {W.}~\bibnamefont
  {Tittel}},\ }\href {http://dx.doi.org/10.1038/nphoton.2009.231} {\bibfield
  {journal} {\bibinfo  {journal} {Nat Photon}\ }\textbf {\bibinfo {volume}
  {3}},\ \bibinfo {pages} {706} (\bibinfo {year} {2009})}\BibitemShut {NoStop}%
\bibitem [{\citenamefont {Sangouard}\ \emph {et~al.}(2011)\citenamefont
  {Sangouard}, \citenamefont {Simon}, \citenamefont {De~Riedmatten},\ and\
  \citenamefont {Gisin}}]{Sangouard:2011bp}%
  \BibitemOpen
  \bibfield  {author} {\bibinfo {author} {\bibfnamefont {N.}~\bibnamefont
  {Sangouard}}, \bibinfo {author} {\bibfnamefont {C.}~\bibnamefont {Simon}},
  \bibinfo {author} {\bibfnamefont {H.}~\bibnamefont {De~Riedmatten}}, \ and\
  \bibinfo {author} {\bibfnamefont {N.}~\bibnamefont {Gisin}},\ }\href
  {\doibase 10.1103/RevModPhys.83.33} {\bibfield  {journal} {\bibinfo
  {journal} {Reviews of Modern Physics}\ }\textbf {\bibinfo {volume} {83}},\
  \bibinfo {pages} {33} (\bibinfo {year} {2011})}\BibitemShut {NoStop}%
\bibitem [{\citenamefont {Kimble}(2008)}]{Kimble08}%
  \BibitemOpen
  \bibfield  {author} {\bibinfo {author} {\bibfnamefont {H.~J.}\ \bibnamefont
  {Kimble}},\ }\href {http://dx.doi.org/10.1038/nature07127} {\bibfield
  {journal} {\bibinfo  {journal} {Nature}\ }\textbf {\bibinfo {volume} {453}},\
  \bibinfo {pages} {1023} (\bibinfo {year} {2008})}\BibitemShut {NoStop}%
\bibitem [{\citenamefont {Knill}\ \emph {et~al.}(2001)\citenamefont {Knill},
  \citenamefont {Laflamme},\ and\ \citenamefont {Milburn}}]{Knill01}%
  \BibitemOpen
  \bibfield  {author} {\bibinfo {author} {\bibfnamefont {E.}~\bibnamefont
  {Knill}}, \bibinfo {author} {\bibfnamefont {R.}~\bibnamefont {Laflamme}}, \
  and\ \bibinfo {author} {\bibfnamefont {G.~J.}\ \bibnamefont {Milburn}},\
  }\href {http://dx.doi.org/10.1038/35051009} {\bibfield  {journal} {\bibinfo
  {journal} {Nature}\ }\textbf {\bibinfo {volume} {409}},\ \bibinfo {pages}
  {46} (\bibinfo {year} {2001})}\BibitemShut {NoStop}%
\bibitem [{\citenamefont {Chen}\ \emph {et~al.}(2013)\citenamefont {Chen},
  \citenamefont {Lee}, \citenamefont {Wang}, \citenamefont {Du}, \citenamefont
  {Chen}, \citenamefont {Chen},\ and\ \citenamefont {Yu}}]{Chen13}%
  \BibitemOpen
  \bibfield  {author} {\bibinfo {author} {\bibfnamefont {Y.-H.}\ \bibnamefont
  {Chen}}, \bibinfo {author} {\bibfnamefont {M.-J.}\ \bibnamefont {Lee}},
  \bibinfo {author} {\bibfnamefont {I.-C.}\ \bibnamefont {Wang}}, \bibinfo
  {author} {\bibfnamefont {S.}~\bibnamefont {Du}}, \bibinfo {author}
  {\bibfnamefont {Y.-F.}\ \bibnamefont {Chen}}, \bibinfo {author}
  {\bibfnamefont {Y.-C.}\ \bibnamefont {Chen}}, \ and\ \bibinfo {author}
  {\bibfnamefont {I.~A.}\ \bibnamefont {Yu}},\ }\href {\doibase
  10.1103/PhysRevLett.110.083601} {\bibfield  {journal} {\bibinfo  {journal}
  {Phys. Rev. Lett.}\ }\textbf {\bibinfo {volume} {110}},\ \bibinfo {pages}
  {083601} (\bibinfo {year} {2013})}\BibitemShut {NoStop}%
\bibitem [{lif()}]{lifetime}%
  \BibitemOpen
  \href@noop {} {}\bibinfo {note} {Lifetime here is defined as the time taken
  for the efficiency to drop by a factor of $e$.}\BibitemShut {Stop}%
\bibitem [{\citenamefont {Heinze}\ \emph {et~al.}(2013)\citenamefont {Heinze},
  \citenamefont {Hubrich},\ and\ \citenamefont {Halfmann}}]{Heinze:2013dh}%
  \BibitemOpen
  \bibfield  {author} {\bibinfo {author} {\bibfnamefont {G.}~\bibnamefont
  {Heinze}}, \bibinfo {author} {\bibfnamefont {C.}~\bibnamefont {Hubrich}}, \
  and\ \bibinfo {author} {\bibfnamefont {T.}~\bibnamefont {Halfmann}},\ }\href
  {\doibase 10.1103/PhysRevLett.111.033601} {\bibfield  {journal} {\bibinfo
  {journal} {Physical Review Letters}\ }\textbf {\bibinfo {volume} {111}},\
  \bibinfo {pages} {033601} (\bibinfo {year} {2013})}\BibitemShut {NoStop}%
\bibitem [{\citenamefont {Reim}\ \emph {et~al.}(2010)\citenamefont {Reim},
  \citenamefont {Nunn}, \citenamefont {Lorenz}, \citenamefont {Sussman},
  \citenamefont {Lee}, \citenamefont {Langford}, \citenamefont {Jaksch},\ and\
  \citenamefont {Walmsley}}]{Reim10}%
  \BibitemOpen
  \bibfield  {author} {\bibinfo {author} {\bibfnamefont {K.~F.}\ \bibnamefont
  {Reim}}, \bibinfo {author} {\bibfnamefont {J.}~\bibnamefont {Nunn}}, \bibinfo
  {author} {\bibfnamefont {V.~O.}\ \bibnamefont {Lorenz}}, \bibinfo {author}
  {\bibfnamefont {B.~J.}\ \bibnamefont {Sussman}}, \bibinfo {author}
  {\bibfnamefont {K.~C.}\ \bibnamefont {Lee}}, \bibinfo {author} {\bibfnamefont
  {N.~K.}\ \bibnamefont {Langford}}, \bibinfo {author} {\bibfnamefont
  {D.}~\bibnamefont {Jaksch}}, \ and\ \bibinfo {author} {\bibfnamefont {I.~A.}\
  \bibnamefont {Walmsley}},\ }\href {http://dx.doi.org/10.1038/nphoton.2010.30}
  {\bibfield  {journal} {\bibinfo  {journal} {Nat Photon}\ }\textbf {\bibinfo
  {volume} {4}},\ \bibinfo {pages} {218} (\bibinfo {year} {2010})}\BibitemShut
  {NoStop}%
\bibitem [{\citenamefont {Reim}\ \emph {et~al.}(2011)\citenamefont {Reim},
  \citenamefont {Michelberger}, \citenamefont {Lee}, \citenamefont {Nunn},
  \citenamefont {Langford},\ and\ \citenamefont {Walmsley}}]{Reim11}%
  \BibitemOpen
  \bibfield  {author} {\bibinfo {author} {\bibfnamefont {K.~F.}\ \bibnamefont
  {Reim}}, \bibinfo {author} {\bibfnamefont {P.}~\bibnamefont {Michelberger}},
  \bibinfo {author} {\bibfnamefont {K.~C.}\ \bibnamefont {Lee}}, \bibinfo
  {author} {\bibfnamefont {J.}~\bibnamefont {Nunn}}, \bibinfo {author}
  {\bibfnamefont {N.~K.}\ \bibnamefont {Langford}}, \ and\ \bibinfo {author}
  {\bibfnamefont {I.~A.}\ \bibnamefont {Walmsley}},\ }\href {\doibase
  10.1103/PhysRevLett.107.053603} {\bibfield  {journal} {\bibinfo  {journal}
  {Phys. Rev. Lett.}\ }\textbf {\bibinfo {volume} {107}},\ \bibinfo {pages}
  {053603} (\bibinfo {year} {2011})}\BibitemShut {NoStop}%
\bibitem [{\citenamefont {Sabooni}\ \emph {et~al.}(2013)\citenamefont
  {Sabooni}, \citenamefont {Li}, \citenamefont {Kroll},\ and\ \citenamefont
  {Rippe}}]{Sabooni:2013cr}%
  \BibitemOpen
  \bibfield  {author} {\bibinfo {author} {\bibfnamefont {M.}~\bibnamefont
  {Sabooni}}, \bibinfo {author} {\bibfnamefont {Q.}~\bibnamefont {Li}},
  \bibinfo {author} {\bibfnamefont {S.}~\bibnamefont {Kroll}}, \ and\ \bibinfo
  {author} {\bibfnamefont {L.}~\bibnamefont {Rippe}},\ }\href {\doibase
  10.1103/PhysRevLett.110.133604} {\bibfield  {journal} {\bibinfo  {journal}
  {Physical Review Letters}\ }\textbf {\bibinfo {volume} {110}},\ \bibinfo
  {pages} {133604} (\bibinfo {year} {2013})}\BibitemShut {NoStop}%
\bibitem [{\citenamefont {Jobez}\ \emph {et~al.}(2015)\citenamefont {Jobez},
  \citenamefont {Laplane}, \citenamefont {Timoney}, \citenamefont {Gisin},
  \citenamefont {Ferrier}, \citenamefont {Goldner},\ and\ \citenamefont
  {Afzelius}}]{Jobez:2015gt}%
  \BibitemOpen
  \bibfield  {author} {\bibinfo {author} {\bibfnamefont {P.}~\bibnamefont
  {Jobez}}, \bibinfo {author} {\bibfnamefont {C.}~\bibnamefont {Laplane}},
  \bibinfo {author} {\bibfnamefont {N.}~\bibnamefont {Timoney}}, \bibinfo
  {author} {\bibfnamefont {N.}~\bibnamefont {Gisin}}, \bibinfo {author}
  {\bibfnamefont {A.}~\bibnamefont {Ferrier}}, \bibinfo {author} {\bibfnamefont
  {P.}~\bibnamefont {Goldner}}, \ and\ \bibinfo {author} {\bibfnamefont
  {M.}~\bibnamefont {Afzelius}},\ }\href {\doibase
  10.1103/PhysRevLett.114.230502} {\bibfield  {journal} {\bibinfo  {journal}
  {Physical Review Letters}\ }\textbf {\bibinfo {volume} {114}},\ \bibinfo
  {pages} {230502} (\bibinfo {year} {2015})}\BibitemShut {NoStop}%
\bibitem [{\citenamefont {Hosseini}\ \emph
  {et~al.}(2011{\natexlab{a}})\citenamefont {Hosseini}, \citenamefont
  {Sparkes}, \citenamefont {Campbell}, \citenamefont {Lam},\ and\ \citenamefont
  {Buchler}}]{Hosseini11}%
  \BibitemOpen
  \bibfield  {author} {\bibinfo {author} {\bibfnamefont {M.}~\bibnamefont
  {Hosseini}}, \bibinfo {author} {\bibfnamefont {B.~M.}\ \bibnamefont
  {Sparkes}}, \bibinfo {author} {\bibfnamefont {G.}~\bibnamefont {Campbell}},
  \bibinfo {author} {\bibfnamefont {P.~K.}\ \bibnamefont {Lam}}, \ and\
  \bibinfo {author} {\bibfnamefont {B.~C.}\ \bibnamefont {Buchler}},\ }\href
  {http://dx.doi.org/10.1038/ncomms1175} {\bibfield  {journal} {\bibinfo
  {journal} {Nat Commun}\ }\textbf {\bibinfo {volume} {2}},\ \bibinfo {pages}
  {174} (\bibinfo {year} {2011}{\natexlab{a}})}\BibitemShut {NoStop}%
\bibitem [{\citenamefont {Hedges}\ \emph {et~al.}(2010)\citenamefont {Hedges},
  \citenamefont {Longdell}, \citenamefont {Li},\ and\ \citenamefont
  {Sellars}}]{Hedges10}%
  \BibitemOpen
  \bibfield  {author} {\bibinfo {author} {\bibfnamefont {M.~P.}\ \bibnamefont
  {Hedges}}, \bibinfo {author} {\bibfnamefont {J.~J.}\ \bibnamefont
  {Longdell}}, \bibinfo {author} {\bibfnamefont {Y.}~\bibnamefont {Li}}, \ and\
  \bibinfo {author} {\bibfnamefont {M.~J.}\ \bibnamefont {Sellars}},\ }\href
  {http://dx.doi.org/10.1038/nature09081} {\bibfield  {journal} {\bibinfo
  {journal} {Nature}\ }\textbf {\bibinfo {volume} {465}},\ \bibinfo {pages}
  {1052} (\bibinfo {year} {2010})}\BibitemShut {NoStop}%
\bibitem [{\citenamefont {Sparkes}\ \emph {et~al.}(2013)\citenamefont
  {Sparkes}, \citenamefont {Bernu}, \citenamefont {Hosseini}, \citenamefont
  {Geng}, \citenamefont {Glorieux}, \citenamefont {Altin}, \citenamefont {Lam},
  \citenamefont {Robins},\ and\ \citenamefont {Buchler}}]{Sparkes13}%
  \BibitemOpen
  \bibfield  {author} {\bibinfo {author} {\bibfnamefont {B.~M.}\ \bibnamefont
  {Sparkes}}, \bibinfo {author} {\bibfnamefont {J.}~\bibnamefont {Bernu}},
  \bibinfo {author} {\bibfnamefont {M.}~\bibnamefont {Hosseini}}, \bibinfo
  {author} {\bibfnamefont {J.}~\bibnamefont {Geng}}, \bibinfo {author}
  {\bibfnamefont {Q.}~\bibnamefont {Glorieux}}, \bibinfo {author}
  {\bibfnamefont {P.~A.}\ \bibnamefont {Altin}}, \bibinfo {author}
  {\bibfnamefont {P.~K.}\ \bibnamefont {Lam}}, \bibinfo {author} {\bibfnamefont
  {N.~P.}\ \bibnamefont {Robins}}, \ and\ \bibinfo {author} {\bibfnamefont
  {B.~C.}\ \bibnamefont {Buchler}},\ }\href
  {http://stacks.iop.org/1367-2630/15/i=8/a=085027} {\bibfield  {journal}
  {\bibinfo  {journal} {New Journal of Physics}\ }\textbf {\bibinfo {volume}
  {15}},\ \bibinfo {pages} {085027} (\bibinfo {year} {2013})}\BibitemShut
  {NoStop}%
\bibitem [{\citenamefont {Grosshans}\ and\ \citenamefont
  {Grangier}(2001)}]{Grosshans01}%
  \BibitemOpen
  \bibfield  {author} {\bibinfo {author} {\bibfnamefont {F.}~\bibnamefont
  {Grosshans}}\ and\ \bibinfo {author} {\bibfnamefont {P.}~\bibnamefont
  {Grangier}},\ }\href {http://link.aps.org/doi/10.1103/PhysRevA.64.010301}
  {\bibfield  {journal} {\bibinfo  {journal} {Physical Review A}\ }\textbf
  {\bibinfo {volume} {64}},\ \bibinfo {pages} {010301} (\bibinfo {year}
  {2001})}\BibitemShut {NoStop}%
\bibitem [{\citenamefont {Varnava}\ \emph {et~al.}(2006)\citenamefont
  {Varnava}, \citenamefont {Browne},\ and\ \citenamefont
  {Rudolph}}]{Varnava06}%
  \BibitemOpen
  \bibfield  {author} {\bibinfo {author} {\bibfnamefont {M.}~\bibnamefont
  {Varnava}}, \bibinfo {author} {\bibfnamefont {D.~E.}\ \bibnamefont {Browne}},
  \ and\ \bibinfo {author} {\bibfnamefont {T.}~\bibnamefont {Rudolph}},\ }\href
  {http://link.aps.org/doi/10.1103/PhysRevLett.97.120501} {\bibfield  {journal}
  {\bibinfo  {journal} {Physical Review Letters}\ }\textbf {\bibinfo {volume}
  {97}},\ \bibinfo {pages} {120501} (\bibinfo {year} {2006})}\BibitemShut
  {NoStop}%
\bibitem [{\citenamefont {Hosseini}\ \emph
  {et~al.}(2011{\natexlab{b}})\citenamefont {Hosseini}, \citenamefont
  {Campbell}, \citenamefont {Sparkes}, \citenamefont {Lam},\ and\ \citenamefont
  {Buchler}}]{Hosseini11a}%
  \BibitemOpen
  \bibfield  {author} {\bibinfo {author} {\bibfnamefont {M.}~\bibnamefont
  {Hosseini}}, \bibinfo {author} {\bibfnamefont {G.}~\bibnamefont {Campbell}},
  \bibinfo {author} {\bibfnamefont {B.~M.}\ \bibnamefont {Sparkes}}, \bibinfo
  {author} {\bibfnamefont {P.~K.}\ \bibnamefont {Lam}}, \ and\ \bibinfo
  {author} {\bibfnamefont {B.~C.}\ \bibnamefont {Buchler}},\ }\href
  {http://dx.doi.org/10.1038/nphys2021} {\bibfield  {journal} {\bibinfo
  {journal} {Nat Phys}\ }\textbf {\bibinfo {volume} {7}},\ \bibinfo {pages}
  {794} (\bibinfo {year} {2011}{\natexlab{b}})}\BibitemShut {NoStop}%
\bibitem [{\citenamefont {Riedl}\ \emph {et~al.}(2012)\citenamefont {Riedl},
  \citenamefont {Lettner}, \citenamefont {Vo}, \citenamefont {Baur},
  \citenamefont {Rempe},\ and\ \citenamefont {D{\"u}rr}}]{Riedl:2012gv}%
  \BibitemOpen
  \bibfield  {author} {\bibinfo {author} {\bibfnamefont {S.}~\bibnamefont
  {Riedl}}, \bibinfo {author} {\bibfnamefont {M.}~\bibnamefont {Lettner}},
  \bibinfo {author} {\bibfnamefont {C.}~\bibnamefont {Vo}}, \bibinfo {author}
  {\bibfnamefont {S.}~\bibnamefont {Baur}}, \bibinfo {author} {\bibfnamefont
  {G.}~\bibnamefont {Rempe}}, \ and\ \bibinfo {author} {\bibfnamefont
  {S.}~\bibnamefont {D{\"u}rr}},\ }\href {\doibase 10.1103/PhysRevA.85.022318}
  {\bibfield  {journal} {\bibinfo  {journal} {Physical Review A}\ }\textbf
  {\bibinfo {volume} {85}},\ \bibinfo {pages} {022318} (\bibinfo {year}
  {2012})}\BibitemShut {NoStop}%
\bibitem [{\citenamefont {Jobez}\ \emph {et~al.}(2014)\citenamefont {Jobez},
  \citenamefont {Usmani}, \citenamefont {Timoney}, \citenamefont {Laplane},
  \citenamefont {Gisin},\ and\ \citenamefont {Afzelius}}]{Afzelius:2014ua}%
  \BibitemOpen
  \bibfield  {author} {\bibinfo {author} {\bibfnamefont {P.}~\bibnamefont
  {Jobez}}, \bibinfo {author} {\bibfnamefont {I.}~\bibnamefont {Usmani}},
  \bibinfo {author} {\bibfnamefont {N.}~\bibnamefont {Timoney}}, \bibinfo
  {author} {\bibfnamefont {C.}~\bibnamefont {Laplane}}, \bibinfo {author}
  {\bibfnamefont {N.}~\bibnamefont {Gisin}}, \ and\ \bibinfo {author}
  {\bibfnamefont {M.}~\bibnamefont {Afzelius}},\ }\href
  {http://www.scopus.com/scopus/record/display.url?fedsrfIntegrator=MEKPAPERS-SCOCIT&origin=fedsrf&view=basic&eid=2-s2.0-84907346554}
  {\bibfield  {journal} {\bibinfo  {journal} {New Journal of Physics}\ }\textbf
  {\bibinfo {volume} {16}},\ \bibinfo {pages} {083005} (\bibinfo {year}
  {2014})}\BibitemShut {NoStop}%
\bibitem [{\citenamefont {Geng}\ \emph {et~al.}(2014)\citenamefont {Geng},
  \citenamefont {Campbell}, \citenamefont {Bernu}, \citenamefont
  {Higginbottom}, \citenamefont {Sparkes}, \citenamefont {Assad}, \citenamefont
  {Zhang}, \citenamefont {Robins}, \citenamefont {Lam},\ and\ \citenamefont
  {Buchler}}]{Geng14}%
  \BibitemOpen
  \bibfield  {author} {\bibinfo {author} {\bibfnamefont {J.}~\bibnamefont
  {Geng}}, \bibinfo {author} {\bibfnamefont {G.~T.}\ \bibnamefont {Campbell}},
  \bibinfo {author} {\bibfnamefont {J.}~\bibnamefont {Bernu}}, \bibinfo
  {author} {\bibfnamefont {D.~B.}\ \bibnamefont {Higginbottom}}, \bibinfo
  {author} {\bibfnamefont {B.~M.}\ \bibnamefont {Sparkes}}, \bibinfo {author}
  {\bibfnamefont {S.~M.}\ \bibnamefont {Assad}}, \bibinfo {author}
  {\bibfnamefont {W.~P.}\ \bibnamefont {Zhang}}, \bibinfo {author}
  {\bibfnamefont {N.~P.}\ \bibnamefont {Robins}}, \bibinfo {author}
  {\bibfnamefont {P.~K.}\ \bibnamefont {Lam}}, \ and\ \bibinfo {author}
  {\bibfnamefont {B.~C.}\ \bibnamefont {Buchler}},\ }\href
  {http://stacks.iop.org/1367-2630/16/i=11/a=113053} {\bibfield  {journal}
  {\bibinfo  {journal} {New Journal of Physics}\ }\textbf {\bibinfo {volume}
  {16}},\ \bibinfo {pages} {113053} (\bibinfo {year} {2014})}\BibitemShut
  {NoStop}%
\bibitem [{\citenamefont {Petrich}\ \emph {et~al.}(1994)\citenamefont
  {Petrich}, \citenamefont {Anderson}, \citenamefont {Ensher},\ and\
  \citenamefont {Cornell}}]{Petrich94}%
  \BibitemOpen
  \bibfield  {author} {\bibinfo {author} {\bibfnamefont {W.}~\bibnamefont
  {Petrich}}, \bibinfo {author} {\bibfnamefont {M.~H.}\ \bibnamefont
  {Anderson}}, \bibinfo {author} {\bibfnamefont {J.~R.}\ \bibnamefont
  {Ensher}}, \ and\ \bibinfo {author} {\bibfnamefont {E.~A.}\ \bibnamefont
  {Cornell}},\ }\href@noop {} {\bibfield  {journal} {\bibinfo  {journal}
  {Journal of the Optical Society of America B}\ }\textbf {\bibinfo {volume}
  {11}},\ \bibinfo {pages} {1332} (\bibinfo {year} {1994})}\BibitemShut
  {NoStop}%
\bibitem [{\citenamefont {Hsiao}\ \emph {et~al.}(2014)\citenamefont {Hsiao},
  \citenamefont {Chen}, \citenamefont {Tsai},\ and\ \citenamefont
  {Chen}}]{Hsiao14}%
  \BibitemOpen
  \bibfield  {author} {\bibinfo {author} {\bibfnamefont {Y.-F.}\ \bibnamefont
  {Hsiao}}, \bibinfo {author} {\bibfnamefont {H.-S.}\ \bibnamefont {Chen}},
  \bibinfo {author} {\bibfnamefont {P.-J.}\ \bibnamefont {Tsai}}, \ and\
  \bibinfo {author} {\bibfnamefont {Y.-C.}\ \bibnamefont {Chen}},\ }\href
  {http://link.aps.org/doi/10.1103/PhysRevA.90.055401} {\bibfield  {journal}
  {\bibinfo  {journal} {Physical Review A}\ }\textbf {\bibinfo {volume} {90}},\
  \bibinfo {pages} {055401} (\bibinfo {year} {2014})}\BibitemShut {NoStop}%
\bibitem [{sup()}]{supple}%
  \BibitemOpen
  \href@noop {} {\bibinfo  {journal} {See Supplemental Material for details of
  numerical simulations, decay model and multimode storage}\ }\BibitemShut
  {NoStop}%
\bibitem [{\citenamefont {Bussi{\`e}res}\ \emph {et~al.}(2013)\citenamefont
  {Bussi{\`e}res}, \citenamefont {Sangouard}, \citenamefont {Afzelius},
  \citenamefont {de~Riedmatten}, \citenamefont {Simon},\ and\ \citenamefont
  {Tittel}}]{bussieres2013prospective}%
  \BibitemOpen
\bibfield  {journal} {  }\bibfield  {author} {\bibinfo {author} {\bibfnamefont
  {F.}~\bibnamefont {Bussi{\`e}res}}, \bibinfo {author} {\bibfnamefont
  {N.}~\bibnamefont {Sangouard}}, \bibinfo {author} {\bibfnamefont
  {M.}~\bibnamefont {Afzelius}}, \bibinfo {author} {\bibfnamefont
  {H.}~\bibnamefont {de~Riedmatten}}, \bibinfo {author} {\bibfnamefont
  {C.}~\bibnamefont {Simon}}, \ and\ \bibinfo {author} {\bibfnamefont
  {W.}~\bibnamefont {Tittel}},\ }\href@noop {} {\bibfield  {journal} {\bibinfo
  {journal} {Journal of Modern Optics}\ }\textbf {\bibinfo {volume} {60}},\
  \bibinfo {pages} {1519} (\bibinfo {year} {2013})}\BibitemShut {NoStop}%
\bibitem [{\citenamefont {Jenkins}\ \emph {et~al.}(2012)\citenamefont
  {Jenkins}, \citenamefont {Zhang},\ and\ \citenamefont {Kennedy}}]{Jenkins12}%
  \BibitemOpen
  \bibfield  {author} {\bibinfo {author} {\bibfnamefont {S.~D.}\ \bibnamefont
  {Jenkins}}, \bibinfo {author} {\bibfnamefont {T.}~\bibnamefont {Zhang}}, \
  and\ \bibinfo {author} {\bibfnamefont {T.~A.~B.}\ \bibnamefont {Kennedy}},\
  }\href {http://stacks.iop.org/0953-4075/45/i=12/a=124005} {\bibfield
  {journal} {\bibinfo  {journal} {Journal of Physics B: Atomic, Molecular and
  Optical Physics}\ }\textbf {\bibinfo {volume} {45}},\ \bibinfo {pages}
  {124005} (\bibinfo {year} {2012})}\BibitemShut {NoStop}%
\bibitem [{\citenamefont {H{\'e}tet}\ \emph {et~al.}(2008)\citenamefont
  {H{\'e}tet}, \citenamefont {Longdell}, \citenamefont {Sellars}, \citenamefont
  {Lam},\ and\ \citenamefont {Buchler}}]{Hetet08}%
  \BibitemOpen
  \bibfield  {author} {\bibinfo {author} {\bibfnamefont {G.}~\bibnamefont
  {H{\'e}tet}}, \bibinfo {author} {\bibfnamefont {J.~J.}\ \bibnamefont
  {Longdell}}, \bibinfo {author} {\bibfnamefont {M.~J.}\ \bibnamefont
  {Sellars}}, \bibinfo {author} {\bibfnamefont {P.~K.}\ \bibnamefont {Lam}}, \
  and\ \bibinfo {author} {\bibfnamefont {B.~C.}\ \bibnamefont {Buchler}},\
  }\href {http://link.aps.org/doi/10.1103/PhysRevLett.101.203601} {\bibfield
  {journal} {\bibinfo  {journal} {Physical Review Letters}\ }\textbf {\bibinfo
  {volume} {101}},\ \bibinfo {pages} {203601} (\bibinfo {year}
  {2008})}\BibitemShut {NoStop}%
\bibitem [{\citenamefont {Luo}\ \emph {et~al.}(2013)\citenamefont {Luo},
  \citenamefont {Hope}, \citenamefont {Hillman},\ and\ \citenamefont
  {Stace}}]{Luo13}%
  \BibitemOpen
  \bibfield  {author} {\bibinfo {author} {\bibfnamefont {X.~W.}\ \bibnamefont
  {Luo}}, \bibinfo {author} {\bibfnamefont {J.~J.}\ \bibnamefont {Hope}},
  \bibinfo {author} {\bibfnamefont {B.}~\bibnamefont {Hillman}}, \ and\
  \bibinfo {author} {\bibfnamefont {T.~M.}\ \bibnamefont {Stace}},\ }\href
  {http://link.aps.org/doi/10.1103/PhysRevA.87.062328} {\bibfield  {journal}
  {\bibinfo  {journal} {Physical Review A}\ }\textbf {\bibinfo {volume} {87}},\
  \bibinfo {pages} {062328} (\bibinfo {year} {2013})}\BibitemShut {NoStop}%
\bibitem [{\citenamefont {Bao}\ \emph {et~al.}(2012)\citenamefont {Bao},
  \citenamefont {Reingruber}, \citenamefont {Dietrich}, \citenamefont {Rui},
  \citenamefont {Duck}, \citenamefont {Strassel}, \citenamefont {Li},
  \citenamefont {Liu}, \citenamefont {Zhao},\ and\ \citenamefont
  {Pan}}]{Bao12}%
  \BibitemOpen
  \bibfield  {author} {\bibinfo {author} {\bibfnamefont {X.-H.}\ \bibnamefont
  {Bao}}, \bibinfo {author} {\bibfnamefont {A.}~\bibnamefont {Reingruber}},
  \bibinfo {author} {\bibfnamefont {P.}~\bibnamefont {Dietrich}}, \bibinfo
  {author} {\bibfnamefont {J.}~\bibnamefont {Rui}}, \bibinfo {author}
  {\bibfnamefont {A.}~\bibnamefont {Duck}}, \bibinfo {author} {\bibfnamefont
  {T.}~\bibnamefont {Strassel}}, \bibinfo {author} {\bibfnamefont
  {L.}~\bibnamefont {Li}}, \bibinfo {author} {\bibfnamefont {N.-L.}\
  \bibnamefont {Liu}}, \bibinfo {author} {\bibfnamefont {B.}~\bibnamefont
  {Zhao}}, \ and\ \bibinfo {author} {\bibfnamefont {J.-W.}\ \bibnamefont
  {Pan}},\ }\href {http://dx.doi.org/10.1038/nphys2324} {\bibfield  {journal}
  {\bibinfo  {journal} {Nat Phys}\ }\textbf {\bibinfo {volume} {8}},\ \bibinfo
  {pages} {517} (\bibinfo {year} {2012})}\BibitemShut {NoStop}%
\bibitem [{\citenamefont {Leonhardt}\ and\ \citenamefont
  {Paul}(1995)}]{Leonhardt199589}%
  \BibitemOpen
  \bibfield  {author} {\bibinfo {author} {\bibfnamefont {U.}~\bibnamefont
  {Leonhardt}}\ and\ \bibinfo {author} {\bibfnamefont {H.}~\bibnamefont
  {Paul}},\ }\href {\doibase http://dx.doi.org/10.1016/0079-6727(94)00007-L}
  {\bibfield  {journal} {\bibinfo  {journal} {Progress in Quantum Electronics}\
  }\textbf {\bibinfo {volume} {19}},\ \bibinfo {pages} {89 } (\bibinfo {year}
  {1995})}\BibitemShut {NoStop}%
\bibitem [{\citenamefont {Hetet}\ \emph {et~al.}(2008)\citenamefont {Hetet},
  \citenamefont {Peng}, \citenamefont {Johnsson}, \citenamefont {Hope},\ and\
  \citenamefont {Lam}}]{hetet2008characterization}%
  \BibitemOpen
  \bibfield  {author} {\bibinfo {author} {\bibfnamefont {G.}~\bibnamefont
  {Hetet}}, \bibinfo {author} {\bibfnamefont {A.}~\bibnamefont {Peng}},
  \bibinfo {author} {\bibfnamefont {M.}~\bibnamefont {Johnsson}}, \bibinfo
  {author} {\bibfnamefont {J.}~\bibnamefont {Hope}}, \ and\ \bibinfo {author}
  {\bibfnamefont {P.~K.}\ \bibnamefont {Lam}},\ }\href@noop {} {\bibfield
  {journal} {\bibinfo  {journal} {Physical Review A}\ }\textbf {\bibinfo
  {volume} {77}},\ \bibinfo {pages} {012323} (\bibinfo {year}
  {2008})}\BibitemShut {NoStop}%
\bibitem [{\citenamefont {Ralph}\ and\ \citenamefont {Lam}(1998)}]{Ralph98}%
  \BibitemOpen
  \bibfield  {author} {\bibinfo {author} {\bibfnamefont {T.~C.}\ \bibnamefont
  {Ralph}}\ and\ \bibinfo {author} {\bibfnamefont {P.~K.}\ \bibnamefont
  {Lam}},\ }\href {\doibase 10.1103/PhysRevLett.81.5668} {\bibfield  {journal}
  {\bibinfo  {journal} {Phys. Rev. Lett.}\ }\textbf {\bibinfo {volume} {81}},\
  \bibinfo {pages} {5668} (\bibinfo {year} {1998})}\BibitemShut {NoStop}%
\bibitem [{\citenamefont {Liu}\ \emph {et~al.}(2001)\citenamefont {Liu},
  \citenamefont {Dutton}, \citenamefont {Behroozi},\ and\ \citenamefont
  {Hau}}]{Liu:2001wy}%
  \BibitemOpen
  \bibfield  {author} {\bibinfo {author} {\bibfnamefont {C.}~\bibnamefont
  {Liu}}, \bibinfo {author} {\bibfnamefont {Z.}~\bibnamefont {Dutton}},
  \bibinfo {author} {\bibfnamefont {C.~H.}\ \bibnamefont {Behroozi}}, \ and\
  \bibinfo {author} {\bibfnamefont {L.~V.}\ \bibnamefont {Hau}},\ }\href
  {\doibase 10.1038/35054017} {\bibfield  {journal} {\bibinfo  {journal}
  {Nature}\ }\textbf {\bibinfo {volume} {409}},\ \bibinfo {pages} {490}
  (\bibinfo {year} {2001})}\BibitemShut {NoStop}%
\bibitem [{\citenamefont {Longdell}\ \emph {et~al.}(2005)\citenamefont
  {Longdell}, \citenamefont {Fraval}, \citenamefont {Sellars},\ and\
  \citenamefont {Manson}}]{Longdell:2005ik}%
  \BibitemOpen
  \bibfield  {author} {\bibinfo {author} {\bibfnamefont {J.~J.}\ \bibnamefont
  {Longdell}}, \bibinfo {author} {\bibfnamefont {E.}~\bibnamefont {Fraval}},
  \bibinfo {author} {\bibfnamefont {M.~J.}\ \bibnamefont {Sellars}}, \ and\
  \bibinfo {author} {\bibfnamefont {N.~B.}\ \bibnamefont {Manson}},\ }\href
  {\doibase 10.1103/PhysRevLett.95.063601} {\bibfield  {journal} {\bibinfo
  {journal} {Physical Review Letters}\ }\textbf {\bibinfo {volume} {95}},\
  \bibinfo {pages} {063601} (\bibinfo {year} {2005})}\BibitemShut {NoStop}%
\bibitem [{\citenamefont {Choi}\ \emph {et~al.}(2008)\citenamefont {Choi},
  \citenamefont {Deng}, \citenamefont {Laurat},\ and\ \citenamefont
  {Kimble}}]{Kimble_nature_sph-QM1}%
  \BibitemOpen
  \bibfield  {author} {\bibinfo {author} {\bibfnamefont {K.~S.}\ \bibnamefont
  {Choi}}, \bibinfo {author} {\bibfnamefont {H.}~\bibnamefont {Deng}}, \bibinfo
  {author} {\bibfnamefont {J.}~\bibnamefont {Laurat}}, \ and\ \bibinfo {author}
  {\bibfnamefont {H.~J.}\ \bibnamefont {Kimble}},\ }\href {\doibase
  doi:10.1038/nature06670} {\bibfield  {journal} {\bibinfo  {journal} {Nature}\
  }\textbf {\bibinfo {volume} {452}},\ \bibinfo {pages} {67} (\bibinfo {year}
  {2008})}\BibitemShut {NoStop}%
\bibitem [{\citenamefont {Phillips}\ \emph {et~al.}(2008)\citenamefont
  {Phillips}, \citenamefont {Gorshkov},\ and\ \citenamefont
  {Novikova}}]{Phillips:PRA:2008}%
  \BibitemOpen
  \bibfield  {author} {\bibinfo {author} {\bibfnamefont {N.~B.}\ \bibnamefont
  {Phillips}}, \bibinfo {author} {\bibfnamefont {A.~V.}\ \bibnamefont
  {Gorshkov}}, \ and\ \bibinfo {author} {\bibfnamefont {I.}~\bibnamefont
  {Novikova}},\ }\href {\doibase 10.1103/PhysRevA.78.023801} {\bibfield
  {journal} {\bibinfo  {journal} {Physical Review A}\ }\textbf {\bibinfo
  {volume} {78}},\ \bibinfo {pages} {023801} (\bibinfo {year}
  {2008})}\BibitemShut {NoStop}%
\bibitem [{\citenamefont {Schnorrberger}\ \emph {et~al.}(2009)\citenamefont
  {Schnorrberger}, \citenamefont {Thompson}, \citenamefont {Trotzky},
  \citenamefont {Pugatch}, \citenamefont {Davidson}, \citenamefont {Kuhr},\
  and\ \citenamefont {Bloch}}]{Schnorrberger:2009ev}%
  \BibitemOpen
  \bibfield  {author} {\bibinfo {author} {\bibfnamefont {U.}~\bibnamefont
  {Schnorrberger}}, \bibinfo {author} {\bibfnamefont {J.~D.}\ \bibnamefont
  {Thompson}}, \bibinfo {author} {\bibfnamefont {S.}~\bibnamefont {Trotzky}},
  \bibinfo {author} {\bibfnamefont {R.}~\bibnamefont {Pugatch}}, \bibinfo
  {author} {\bibfnamefont {N.}~\bibnamefont {Davidson}}, \bibinfo {author}
  {\bibfnamefont {S.}~\bibnamefont {Kuhr}}, \ and\ \bibinfo {author}
  {\bibfnamefont {I.}~\bibnamefont {Bloch}},\ }\href {\doibase
  10.1103/PhysRevLett.103.033003} {\bibfield  {journal} {\bibinfo  {journal}
  {Physical Review Letters}\ }\textbf {\bibinfo {volume} {103}},\ \bibinfo
  {pages} {033003} (\bibinfo {year} {2009})}\BibitemShut {NoStop}%
\bibitem [{\citenamefont {Zhang}\ \emph {et~al.}(2009)\citenamefont {Zhang},
  \citenamefont {Garner},\ and\ \citenamefont {Hau}}]{Zhang:2009fi}%
  \BibitemOpen
  \bibfield  {author} {\bibinfo {author} {\bibfnamefont {R.}~\bibnamefont
  {Zhang}}, \bibinfo {author} {\bibfnamefont {S.~R.}\ \bibnamefont {Garner}}, \
  and\ \bibinfo {author} {\bibfnamefont {L.~V.}\ \bibnamefont {Hau}},\ }\href
  {\doibase 10.1103/PhysRevLett.103.233602} {\bibfield  {journal} {\bibinfo
  {journal} {Physical Review Letters}\ }\textbf {\bibinfo {volume} {103}},\
  \bibinfo {pages} {233602} (\bibinfo {year} {2009})}\BibitemShut {NoStop}%
\bibitem [{\citenamefont {Zhao}\ \emph {et~al.}(2009)\citenamefont {Zhao},
  \citenamefont {Dudin}, \citenamefont {Jenkins}, \citenamefont {Campbell},
  \citenamefont {Matsukevich}, \citenamefont {Kennedy},\ and\ \citenamefont
  {Kuzmich}}]{Zhao:2009ic}%
  \BibitemOpen
  \bibfield  {author} {\bibinfo {author} {\bibfnamefont {R.}~\bibnamefont
  {Zhao}}, \bibinfo {author} {\bibfnamefont {Y.~O.}\ \bibnamefont {Dudin}},
  \bibinfo {author} {\bibfnamefont {S.~D.}\ \bibnamefont {Jenkins}}, \bibinfo
  {author} {\bibfnamefont {C.~J.}\ \bibnamefont {Campbell}}, \bibinfo {author}
  {\bibfnamefont {D.~N.}\ \bibnamefont {Matsukevich}}, \bibinfo {author}
  {\bibfnamefont {T.~A.~B.}\ \bibnamefont {Kennedy}}, \ and\ \bibinfo {author}
  {\bibfnamefont {A.}~\bibnamefont {Kuzmich}},\ }\href {\doibase
  10.1038/NPHYS1152} {\bibfield  {journal} {\bibinfo  {journal} {Nature
  Physics}\ }\textbf {\bibinfo {volume} {5}},\ \bibinfo {pages} {100} (\bibinfo
  {year} {2009})}\BibitemShut {NoStop}%
\bibitem [{\citenamefont {Zhou}\ \emph {et~al.}(2012)\citenamefont {Zhou},
  \citenamefont {Zhang}, \citenamefont {Liu}, \citenamefont {Chen},
  \citenamefont {Wen}, \citenamefont {Loy}, \citenamefont {Wong},\ and\
  \citenamefont {Du}}]{Zhou:2012fw}%
  \BibitemOpen
  \bibfield  {author} {\bibinfo {author} {\bibfnamefont {S.}~\bibnamefont
  {Zhou}}, \bibinfo {author} {\bibfnamefont {S.}~\bibnamefont {Zhang}},
  \bibinfo {author} {\bibfnamefont {C.}~\bibnamefont {Liu}}, \bibinfo {author}
  {\bibfnamefont {J.~F.}\ \bibnamefont {Chen}}, \bibinfo {author}
  {\bibfnamefont {J.}~\bibnamefont {Wen}}, \bibinfo {author} {\bibfnamefont
  {M.}~\bibnamefont {Loy}}, \bibinfo {author} {\bibfnamefont {G.}~\bibnamefont
  {Wong}}, \ and\ \bibinfo {author} {\bibfnamefont {S.}~\bibnamefont {Du}},\
  }\href {\doibase 10.1364/OE.20.024124} {\bibfield  {journal} {\bibinfo
  {journal} {Optics Express}\ }\textbf {\bibinfo {volume} {20}},\ \bibinfo
  {pages} {24124} (\bibinfo {year} {2012})}\BibitemShut {NoStop}%
\bibitem [{\citenamefont {England}\ \emph {et~al.}(2013)\citenamefont
  {England}, \citenamefont {Bustard}, \citenamefont {Nunn}, \citenamefont
  {Lausten},\ and\ \citenamefont {Sussman}}]{England:2013uf}%
  \BibitemOpen
  \bibfield  {author} {\bibinfo {author} {\bibfnamefont {D.~G.}\ \bibnamefont
  {England}}, \bibinfo {author} {\bibfnamefont {P.~J.}\ \bibnamefont
  {Bustard}}, \bibinfo {author} {\bibfnamefont {J.}~\bibnamefont {Nunn}},
  \bibinfo {author} {\bibfnamefont {R.}~\bibnamefont {Lausten}}, \ and\
  \bibinfo {author} {\bibfnamefont {B.~J.}\ \bibnamefont {Sussman}},\ }\href
  {http://www.scopus.com/scopus/record/display.url?fedsrfIntegrator=MEKPAPERS-SCOCIT&origin=fedsrf&view=basic&eid=2-s2.0-84890291657}
  {\bibfield  {journal} {\bibinfo  {journal} {Physical Review Letters}\
  }\textbf {\bibinfo {volume} {111}},\ \bibinfo {pages} {243601} (\bibinfo
  {year} {2013})}\BibitemShut {NoStop}%
\bibitem [{\citenamefont {G{\"u}ndo{\u g}an}\ \emph {et~al.}(2013)\citenamefont
  {G{\"u}ndo{\u g}an}, \citenamefont {Mazzera}, \citenamefont {Ledingham},
  \citenamefont {Cristiani},\ and\ \citenamefont
  {De~Riedmatten}}]{deRiedmatten:2013vb}%
  \BibitemOpen
  \bibfield  {author} {\bibinfo {author} {\bibfnamefont {M.}~\bibnamefont
  {G{\"u}ndo{\u g}an}}, \bibinfo {author} {\bibfnamefont {M.}~\bibnamefont
  {Mazzera}}, \bibinfo {author} {\bibfnamefont {P.~M.}\ \bibnamefont
  {Ledingham}}, \bibinfo {author} {\bibfnamefont {M.}~\bibnamefont
  {Cristiani}}, \ and\ \bibinfo {author} {\bibfnamefont {H.}~\bibnamefont
  {De~Riedmatten}},\ }\href {http://stacks.iop.org/1367-2630/15/i=4/a=045012}
  {\bibfield  {journal} {\bibinfo  {journal} {New Journal of Physics}\ }\textbf
  {\bibinfo {volume} {15}},\ \bibinfo {pages} {045012} (\bibinfo {year}
  {2013})}\BibitemShut {NoStop}%
\bibitem [{\citenamefont {Xu}\ \emph {et~al.}(2013)\citenamefont {Xu},
  \citenamefont {Wu}, \citenamefont {Tian}, \citenamefont {Chen}, \citenamefont
  {Zhang}, \citenamefont {Yan}, \citenamefont {Li}, \citenamefont {Wang},
  \citenamefont {Xie},\ and\ \citenamefont {Peng}}]{Xu:2013ir}%
  \BibitemOpen
  \bibfield  {author} {\bibinfo {author} {\bibfnamefont {Z.}~\bibnamefont
  {Xu}}, \bibinfo {author} {\bibfnamefont {Y.}~\bibnamefont {Wu}}, \bibinfo
  {author} {\bibfnamefont {L.}~\bibnamefont {Tian}}, \bibinfo {author}
  {\bibfnamefont {L.}~\bibnamefont {Chen}}, \bibinfo {author} {\bibfnamefont
  {Z.}~\bibnamefont {Zhang}}, \bibinfo {author} {\bibfnamefont
  {Z.}~\bibnamefont {Yan}}, \bibinfo {author} {\bibfnamefont {S.}~\bibnamefont
  {Li}}, \bibinfo {author} {\bibfnamefont {H.}~\bibnamefont {Wang}}, \bibinfo
  {author} {\bibfnamefont {C.}~\bibnamefont {Xie}}, \ and\ \bibinfo {author}
  {\bibfnamefont {K.}~\bibnamefont {Peng}},\ }\href {\doibase
  10.1103/PhysRevLett.111.240503} {\bibfield  {journal} {\bibinfo  {journal}
  {Physical Review Letters}\ }\textbf {\bibinfo {volume} {111}},\ \bibinfo
  {pages} {240503} (\bibinfo {year} {2013})}\BibitemShut {NoStop}%
\bibitem [{\citenamefont {Sprague}\ \emph {et~al.}(2014)\citenamefont
  {Sprague}, \citenamefont {Michelberger}, \citenamefont {Champion},
  \citenamefont {England}, \citenamefont {Nunn}, \citenamefont {Jin},
  \citenamefont {Kolthammer}, \citenamefont {Abdolvand}, \citenamefont
  {Russell},\ and\ \citenamefont {Walmsley}}]{Sprague:2014fd}%
  \BibitemOpen
  \bibfield  {author} {\bibinfo {author} {\bibfnamefont {M.~R.}\ \bibnamefont
  {Sprague}}, \bibinfo {author} {\bibfnamefont {P.~S.}\ \bibnamefont
  {Michelberger}}, \bibinfo {author} {\bibfnamefont {T.~F.~M.}\ \bibnamefont
  {Champion}}, \bibinfo {author} {\bibfnamefont {D.~G.}\ \bibnamefont
  {England}}, \bibinfo {author} {\bibfnamefont {J.}~\bibnamefont {Nunn}},
  \bibinfo {author} {\bibfnamefont {X.~M.}\ \bibnamefont {Jin}}, \bibinfo
  {author} {\bibfnamefont {W.~S.}\ \bibnamefont {Kolthammer}}, \bibinfo
  {author} {\bibfnamefont {A.}~\bibnamefont {Abdolvand}}, \bibinfo {author}
  {\bibfnamefont {P.~S.~J.}\ \bibnamefont {Russell}}, \ and\ \bibinfo {author}
  {\bibfnamefont {I.~A.}\ \bibnamefont {Walmsley}},\ }\href {\doibase
  10.1038/nphoton.2014.45} {\bibfield  {journal} {\bibinfo  {journal} {Nature
  Photonics}\ }\textbf {\bibinfo {volume} {8}},\ \bibinfo {pages} {287}
  (\bibinfo {year} {2014})}\BibitemShut {NoStop}%
\bibitem [{\citenamefont {Yoshikawa}\ \emph {et~al.}(2014)\citenamefont
  {Yoshikawa}, \citenamefont {Makino}, \citenamefont {Kurata}, \citenamefont
  {van Loock},\ and\ \citenamefont {Furusawa}}]{Yoshikawa:2014wf}%
  \BibitemOpen
  \bibfield  {author} {\bibinfo {author} {\bibfnamefont {J.~I.}\ \bibnamefont
  {Yoshikawa}}, \bibinfo {author} {\bibfnamefont {K.}~\bibnamefont {Makino}},
  \bibinfo {author} {\bibfnamefont {S.}~\bibnamefont {Kurata}}, \bibinfo
  {author} {\bibfnamefont {P.}~\bibnamefont {van Loock}}, \ and\ \bibinfo
  {author} {\bibfnamefont {A.}~\bibnamefont {Furusawa}},\ }\href
  {http://www.scopus.com/scopus/record/display.url?fedsrfIntegrator=MEKPAPERS-SCOCIT&origin=fedsrf&view=basic&eid=2-s2.0-84892907418}
  {\bibfield  {journal} {\bibinfo  {journal} {Physical Review X}\ }\textbf
  {\bibinfo {volume} {3}},\ \bibinfo {pages} {041028} (\bibinfo {year}
  {2014})}\BibitemShut {NoStop}%
\bibitem [{\citenamefont {Bimbard}\ \emph {et~al.}(2014)\citenamefont
  {Bimbard}, \citenamefont {Boddeda}, \citenamefont {Vitrant}, \citenamefont
  {Grankin}, \citenamefont {Parigi}, \citenamefont {Stanojevic}, \citenamefont
  {Ourjoumtsev},\ and\ \citenamefont {Grangier}}]{Bimbard:2014fy}%
  \BibitemOpen
  \bibfield  {author} {\bibinfo {author} {\bibfnamefont {E.}~\bibnamefont
  {Bimbard}}, \bibinfo {author} {\bibfnamefont {R.}~\bibnamefont {Boddeda}},
  \bibinfo {author} {\bibfnamefont {N.}~\bibnamefont {Vitrant}}, \bibinfo
  {author} {\bibfnamefont {A.}~\bibnamefont {Grankin}}, \bibinfo {author}
  {\bibfnamefont {V.}~\bibnamefont {Parigi}}, \bibinfo {author} {\bibfnamefont
  {J.}~\bibnamefont {Stanojevic}}, \bibinfo {author} {\bibfnamefont
  {A.}~\bibnamefont {Ourjoumtsev}}, \ and\ \bibinfo {author} {\bibfnamefont
  {P.}~\bibnamefont {Grangier}},\ }\href {\doibase
  10.1103/PhysRevLett.112.033601} {\bibfield  {journal} {\bibinfo  {journal}
  {Physical Review Letters}\ }\textbf {\bibinfo {volume} {112}},\ \bibinfo
  {pages} {033601} (\bibinfo {year} {2014})}\BibitemShut {NoStop}%
\bibitem [{\citenamefont {Nicolas}\ \emph {et~al.}(2014)\citenamefont
  {Nicolas}, \citenamefont {Veissier}, \citenamefont {Giner}, \citenamefont
  {Giacobino}, \citenamefont {Maxein},\ and\ \citenamefont
  {Laurat}}]{Nicolas:2014fn}%
  \BibitemOpen
  \bibfield  {author} {\bibinfo {author} {\bibfnamefont {A.}~\bibnamefont
  {Nicolas}}, \bibinfo {author} {\bibfnamefont {L.}~\bibnamefont {Veissier}},
  \bibinfo {author} {\bibfnamefont {L.}~\bibnamefont {Giner}}, \bibinfo
  {author} {\bibfnamefont {E.}~\bibnamefont {Giacobino}}, \bibinfo {author}
  {\bibfnamefont {D.}~\bibnamefont {Maxein}}, \ and\ \bibinfo {author}
  {\bibfnamefont {J.}~\bibnamefont {Laurat}},\ }\href {\doibase
  10.1038/nphoton.2013.355} {\bibfield  {journal} {\bibinfo  {journal} {Nature
  Photonics}\ }\textbf {\bibinfo {volume} {8}},\ \bibinfo {pages} {234}
  (\bibinfo {year} {2014})}\BibitemShut {NoStop}%
\bibitem [{\citenamefont {Ding}\ \emph {et~al.}(2015)\citenamefont {Ding},
  \citenamefont {Zhang}, \citenamefont {Zhou}, \citenamefont {Shi},
  \citenamefont {Xiang}, \citenamefont {Wang}, \citenamefont {Jiang},
  \citenamefont {Shi},\ and\ \citenamefont {Guo}}]{Ding:2015bg}%
  \BibitemOpen
  \bibfield  {author} {\bibinfo {author} {\bibfnamefont {D.-S.}\ \bibnamefont
  {Ding}}, \bibinfo {author} {\bibfnamefont {W.}~\bibnamefont {Zhang}},
  \bibinfo {author} {\bibfnamefont {Z.-Y.}\ \bibnamefont {Zhou}}, \bibinfo
  {author} {\bibfnamefont {S.}~\bibnamefont {Shi}}, \bibinfo {author}
  {\bibfnamefont {G.-Y.}\ \bibnamefont {Xiang}}, \bibinfo {author}
  {\bibfnamefont {X.-S.}\ \bibnamefont {Wang}}, \bibinfo {author}
  {\bibfnamefont {Y.-K.}\ \bibnamefont {Jiang}}, \bibinfo {author}
  {\bibfnamefont {B.-S.}\ \bibnamefont {Shi}}, \ and\ \bibinfo {author}
  {\bibfnamefont {G.-C.}\ \bibnamefont {Guo}},\ }\href {\doibase
  10.1103/PhysRevLett.114.050502} {\bibfield  {journal} {\bibinfo  {journal}
  {Physical Review Letters}\ }\textbf {\bibinfo {volume} {114}},\ \bibinfo
  {pages} {050502} (\bibinfo {year} {2015})}\BibitemShut {NoStop}%
\bibitem [{\citenamefont {Gouraud}\ \emph {et~al.}(2015)\citenamefont
  {Gouraud}, \citenamefont {Maxein}, \citenamefont {Nicolas}, \citenamefont
  {Morin},\ and\ \citenamefont {Laurat}}]{Gouraud:2015je}%
  \BibitemOpen
  \bibfield  {author} {\bibinfo {author} {\bibfnamefont {B.}~\bibnamefont
  {Gouraud}}, \bibinfo {author} {\bibfnamefont {D.}~\bibnamefont {Maxein}},
  \bibinfo {author} {\bibfnamefont {A.}~\bibnamefont {Nicolas}}, \bibinfo
  {author} {\bibfnamefont {O.}~\bibnamefont {Morin}}, \ and\ \bibinfo {author}
  {\bibfnamefont {J.}~\bibnamefont {Laurat}},\ }\href {\doibase
  10.1103/PhysRevLett.114.180503} {\bibfield  {journal} {\bibinfo  {journal}
  {Physical Review Letters}\ }\textbf {\bibinfo {volume} {114}},\ \bibinfo
  {pages} {180503} (\bibinfo {year} {2015})}\BibitemShut {NoStop}%
\bibitem [{\citenamefont {McAuslan}\ \emph {et~al.}(2011)\citenamefont
  {McAuslan}, \citenamefont {Ledingham}, \citenamefont {Naylor}, \citenamefont
  {Beavan}, \citenamefont {Hedges}, \citenamefont {Sellars},\ and\
  \citenamefont {Longdell}}]{McAuslan:2011up}%
  \BibitemOpen
  \bibfield  {author} {\bibinfo {author} {\bibfnamefont {D.~L.}\ \bibnamefont
  {McAuslan}}, \bibinfo {author} {\bibfnamefont {P.~M.}\ \bibnamefont
  {Ledingham}}, \bibinfo {author} {\bibfnamefont {W.~R.}\ \bibnamefont
  {Naylor}}, \bibinfo {author} {\bibfnamefont {S.~E.}\ \bibnamefont {Beavan}},
  \bibinfo {author} {\bibfnamefont {M.~P.}\ \bibnamefont {Hedges}}, \bibinfo
  {author} {\bibfnamefont {M.~J.}\ \bibnamefont {Sellars}}, \ and\ \bibinfo
  {author} {\bibfnamefont {J.~J.}\ \bibnamefont {Longdell}},\ }\href
  {http://www.scopus.com/scopus/record/display.url?fedsrfIntegrator=MEKPAPERS-SCOCIT&origin=fedsrf&view=basic&eid=2-s2.0-80051625547}
  {\bibfield  {journal} {\bibinfo  {journal} {Physical Review A}\ }\textbf
  {\bibinfo {volume} {84}},\ \bibinfo {pages} {022309} (\bibinfo {year}
  {2011})}\BibitemShut {NoStop}%
\bibitem [{\citenamefont {Dajczgewand}\ \emph {et~al.}(2014)\citenamefont
  {Dajczgewand}, \citenamefont {Le~Gou{\"e}t}, \citenamefont
  {Louchet-Chauvet},\ and\ \citenamefont
  {Chaneli{\`e}re}}]{Dajczgewand:2014et}%
  \BibitemOpen
  \bibfield  {author} {\bibinfo {author} {\bibfnamefont {J.}~\bibnamefont
  {Dajczgewand}}, \bibinfo {author} {\bibfnamefont {J.-L.}\ \bibnamefont
  {Le~Gou{\"e}t}}, \bibinfo {author} {\bibfnamefont {A.}~\bibnamefont
  {Louchet-Chauvet}}, \ and\ \bibinfo {author} {\bibfnamefont {T.}~\bibnamefont
  {Chaneli{\`e}re}},\ }\href {\doibase 10.1364/OL.39.002711} {\bibfield
  {journal} {\bibinfo  {journal} {Optics Letters}\ }\textbf {\bibinfo {volume}
  {39}},\ \bibinfo {pages} {2711} (\bibinfo {year} {2014})}\BibitemShut
  {NoStop}%
\bibitem [{\citenamefont {Saglamyurek}\ \emph {et~al.}(2015)\citenamefont
  {Saglamyurek}, \citenamefont {Jin}, \citenamefont {Verma}, \citenamefont
  {Shaw}, \citenamefont {Marsili}, \citenamefont {Nam}, \citenamefont {Oblak},\
  and\ \citenamefont {Tittel}}]{Saglamyurek:2015es}%
  \BibitemOpen
  \bibfield  {author} {\bibinfo {author} {\bibfnamefont {E.}~\bibnamefont
  {Saglamyurek}}, \bibinfo {author} {\bibfnamefont {J.}~\bibnamefont {Jin}},
  \bibinfo {author} {\bibfnamefont {V.~B.}\ \bibnamefont {Verma}}, \bibinfo
  {author} {\bibfnamefont {M.~D.}\ \bibnamefont {Shaw}}, \bibinfo {author}
  {\bibfnamefont {F.}~\bibnamefont {Marsili}}, \bibinfo {author} {\bibfnamefont
  {S.~W.}\ \bibnamefont {Nam}}, \bibinfo {author} {\bibfnamefont
  {D.}~\bibnamefont {Oblak}}, \ and\ \bibinfo {author} {\bibfnamefont
  {W.}~\bibnamefont {Tittel}},\ }\href {\doibase 10.1038/nphoton.2014.311}
  {\bibfield  {journal} {\bibinfo  {journal} {Nature Photonics}\ }\textbf
  {\bibinfo {volume} {9}},\ \bibinfo {pages} {83} (\bibinfo {year}
  {2015})}\BibitemShut {NoStop}%
\bibitem [{\citenamefont {Lauritzen}\ \emph {et~al.}(2010)\citenamefont
  {Lauritzen}, \citenamefont {Min{\'a}{\v r}}, \citenamefont {De~Riedmatten},
  \citenamefont {Afzelius}, \citenamefont {Sangouard}, \citenamefont {Simon},\
  and\ \citenamefont {Gisin}}]{Lauritzen:2010go}%
  \BibitemOpen
  \bibfield  {author} {\bibinfo {author} {\bibfnamefont {B.}~\bibnamefont
  {Lauritzen}}, \bibinfo {author} {\bibfnamefont {J.}~\bibnamefont {Min{\'a}{\v
  r}}}, \bibinfo {author} {\bibfnamefont {H.}~\bibnamefont {De~Riedmatten}},
  \bibinfo {author} {\bibfnamefont {M.}~\bibnamefont {Afzelius}}, \bibinfo
  {author} {\bibfnamefont {N.}~\bibnamefont {Sangouard}}, \bibinfo {author}
  {\bibfnamefont {C.}~\bibnamefont {Simon}}, \ and\ \bibinfo {author}
  {\bibfnamefont {N.}~\bibnamefont {Gisin}},\ }\href {\doibase
  10.1103/PhysRevLett.104.080502} {\bibfield  {journal} {\bibinfo  {journal}
  {Physical Review Letters}\ }\textbf {\bibinfo {volume} {104}},\ \bibinfo
  {pages} {080502} (\bibinfo {year} {2010})}\BibitemShut {NoStop}%
\bibitem [{\citenamefont {Tanzilli}\ \emph {et~al.}(2005)\citenamefont
  {Tanzilli}, \citenamefont {Tittel}, \citenamefont {Halder}, \citenamefont
  {Alibart}, \citenamefont {Baldi}, \citenamefont {Gisin},\ and\ \citenamefont
  {Zbinden}}]{Tanzilli:2005jd}%
  \BibitemOpen
  \bibfield  {author} {\bibinfo {author} {\bibfnamefont {S.}~\bibnamefont
  {Tanzilli}}, \bibinfo {author} {\bibfnamefont {W.}~\bibnamefont {Tittel}},
  \bibinfo {author} {\bibfnamefont {M.}~\bibnamefont {Halder}}, \bibinfo
  {author} {\bibfnamefont {O.}~\bibnamefont {Alibart}}, \bibinfo {author}
  {\bibfnamefont {P.}~\bibnamefont {Baldi}}, \bibinfo {author} {\bibfnamefont
  {N.}~\bibnamefont {Gisin}}, \ and\ \bibinfo {author} {\bibfnamefont
  {H.}~\bibnamefont {Zbinden}},\ }\href {\doibase 10.1038/nature04009}
  {\bibfield  {journal} {\bibinfo  {journal} {Nature}\ }\textbf {\bibinfo
  {volume} {437}},\ \bibinfo {pages} {116} (\bibinfo {year}
  {2005})}\BibitemShut {NoStop}%
\bibitem [{\citenamefont {Ikuta}\ \emph {et~al.}(2011)\citenamefont {Ikuta},
  \citenamefont {Kusaka}, \citenamefont {Kitano}, \citenamefont {Kato},
  \citenamefont {Yamamoto}, \citenamefont {Koashi},\ and\ \citenamefont
  {Imoto}}]{Ikuta:2011de}%
  \BibitemOpen
  \bibfield  {author} {\bibinfo {author} {\bibfnamefont {R.}~\bibnamefont
  {Ikuta}}, \bibinfo {author} {\bibfnamefont {Y.}~\bibnamefont {Kusaka}},
  \bibinfo {author} {\bibfnamefont {T.}~\bibnamefont {Kitano}}, \bibinfo
  {author} {\bibfnamefont {H.}~\bibnamefont {Kato}}, \bibinfo {author}
  {\bibfnamefont {T.}~\bibnamefont {Yamamoto}}, \bibinfo {author}
  {\bibfnamefont {M.}~\bibnamefont {Koashi}}, \ and\ \bibinfo {author}
  {\bibfnamefont {N.}~\bibnamefont {Imoto}},\ }\href {\doibase
  10.1038/ncomms1544} {\bibfield  {journal} {\bibinfo  {journal} {Nature
  Communications}\ }\textbf {\bibinfo {volume} {2}},\ \bibinfo {pages} {1544}
  (\bibinfo {year} {2011})}\BibitemShut {NoStop}%
\bibitem [{\citenamefont {Zaske}\ \emph {et~al.}(2011)\citenamefont {Zaske},
  \citenamefont {Lenhard},\ and\ \citenamefont {Becher}}]{Zaske:2011id}%
  \BibitemOpen
  \bibfield  {author} {\bibinfo {author} {\bibfnamefont {S.}~\bibnamefont
  {Zaske}}, \bibinfo {author} {\bibfnamefont {A.}~\bibnamefont {Lenhard}}, \
  and\ \bibinfo {author} {\bibfnamefont {C.}~\bibnamefont {Becher}},\ }\href
  {\doibase 10.1364/OE.19.012825} {\bibfield  {journal} {\bibinfo  {journal}
  {Optics Express}\ }\textbf {\bibinfo {volume} {19}},\ \bibinfo {pages}
  {12825} (\bibinfo {year} {2011})}\BibitemShut {NoStop}%
\bibitem [{\citenamefont {Albrecht}\ \emph {et~al.}(2014)\citenamefont
  {Albrecht}, \citenamefont {Farrera}, \citenamefont {Fernandez-Gonzalvo},
  \citenamefont {Cristiani},\ and\ \citenamefont
  {De~Riedmatten}}]{Albrecht:2014cf}%
  \BibitemOpen
  \bibfield  {author} {\bibinfo {author} {\bibfnamefont {B.}~\bibnamefont
  {Albrecht}}, \bibinfo {author} {\bibfnamefont {P.}~\bibnamefont {Farrera}},
  \bibinfo {author} {\bibfnamefont {X.}~\bibnamefont {Fernandez-Gonzalvo}},
  \bibinfo {author} {\bibfnamefont {M.}~\bibnamefont {Cristiani}}, \ and\
  \bibinfo {author} {\bibfnamefont {H.}~\bibnamefont {De~Riedmatten}},\ }\href
  {\doibase 10.1038/ncomms4376} {\bibfield  {journal} {\bibinfo  {journal}
  {Nature Communications}\ }\textbf {\bibinfo {volume} {5}},\ \bibinfo {pages}
  {3376} (\bibinfo {year} {2014})}\BibitemShut {NoStop}%
\bibitem [{\citenamefont {Maring}\ \emph {et~al.}(2014)\citenamefont {Maring},
  \citenamefont {Kutluer}, \citenamefont {Cohen}, \citenamefont {Cristiani},
  \citenamefont {Mazzera}, \citenamefont {Ledingham},\ and\ \citenamefont
  {De~Riedmatten}}]{Maring:dp}%
  \BibitemOpen
  \bibfield  {author} {\bibinfo {author} {\bibfnamefont {N.}~\bibnamefont
  {Maring}}, \bibinfo {author} {\bibfnamefont {K.}~\bibnamefont {Kutluer}},
  \bibinfo {author} {\bibfnamefont {J.}~\bibnamefont {Cohen}}, \bibinfo
  {author} {\bibfnamefont {M.}~\bibnamefont {Cristiani}}, \bibinfo {author}
  {\bibfnamefont {M.}~\bibnamefont {Mazzera}}, \bibinfo {author} {\bibfnamefont
  {P.~M.}\ \bibnamefont {Ledingham}}, \ and\ \bibinfo {author} {\bibfnamefont
  {H.}~\bibnamefont {De~Riedmatten}},\ }\href {\doibase
  10.1088/1367-2630/16/11/113021} {\bibfield  {journal} {\bibinfo  {journal}
  {New Journal of Physics}\ }\textbf {\bibinfo {volume} {16}},\ \bibinfo
  {pages} {113021} (\bibinfo {year} {2014})}\BibitemShut {NoStop}%
\bibitem [{\citenamefont {Dudin}\ \emph {et~al.}(2013)\citenamefont {Dudin},
  \citenamefont {Li},\ and\ \citenamefont {Kuzmich}}]{Dudin:2013ew}%
  \BibitemOpen
  \bibfield  {author} {\bibinfo {author} {\bibfnamefont {Y.~O.}\ \bibnamefont
  {Dudin}}, \bibinfo {author} {\bibfnamefont {L.}~\bibnamefont {Li}}, \ and\
  \bibinfo {author} {\bibfnamefont {A.}~\bibnamefont {Kuzmich}},\ }\href
  {\doibase 10.1103/PhysRevA.87.031801} {\bibfield  {journal} {\bibinfo
  {journal} {Physical Review A}\ }\textbf {\bibinfo {volume} {87}},\ \bibinfo
  {pages} {031801} (\bibinfo {year} {2013})}\BibitemShut {NoStop}%
\bibitem [{\citenamefont {Zhong}\ \emph {et~al.}(2015)\citenamefont {Zhong},
  \citenamefont {Hedges}, \citenamefont {Ahlefeldt}, \citenamefont
  {Bartholomew}, \citenamefont {Beavan}, \citenamefont {Wittig}, \citenamefont
  {Longdell},\ and\ \citenamefont {Sellars}}]{Zhong15}%
  \BibitemOpen
  \bibfield  {author} {\bibinfo {author} {\bibfnamefont {M.}~\bibnamefont
  {Zhong}}, \bibinfo {author} {\bibfnamefont {M.~P.}\ \bibnamefont {Hedges}},
  \bibinfo {author} {\bibfnamefont {R.~L.}\ \bibnamefont {Ahlefeldt}}, \bibinfo
  {author} {\bibfnamefont {J.~G.}\ \bibnamefont {Bartholomew}}, \bibinfo
  {author} {\bibfnamefont {S.~E.}\ \bibnamefont {Beavan}}, \bibinfo {author}
  {\bibfnamefont {S.~M.}\ \bibnamefont {Wittig}}, \bibinfo {author}
  {\bibfnamefont {J.~J.}\ \bibnamefont {Longdell}}, \ and\ \bibinfo {author}
  {\bibfnamefont {M.~J.}\ \bibnamefont {Sellars}},\ }\href {\doibase
  10.1038/nature14025} {\bibfield  {journal} {\bibinfo  {journal} {Nature}\
  }\textbf {\bibinfo {volume} {517}},\ \bibinfo {pages} {177} (\bibinfo {year}
  {2015})}\BibitemShut {NoStop}%
\end{thebibliography}%

\end{document}